\documentclass[aps,pra,preprint,groupedaddress,showkeys,floatfix,showpacs]{revtex4}
\usepackage{bm}
\usepackage{setspace}
\usepackage{mathtools}
\usepackage{amsmath}

\begin{document}

\title{Symmetries in elastic scattering of electrons by hydrogen atoms
 in a two-color bicircular laser field
}

\author{Gabriela Buica}
\email{buica@spacescience.ro}
\affiliation{
Institute of Space Sciences, P.O. Box MG-36, Ro 77125,
Bucharest-M\u{a}gurele, Romania}

\begin{abstract}
 We consider the elastic scattering of  electrons by hydrogen atoms
 in the presence of a two-color circularly polarized laser field
 in the domain of moderate intensities  below $10^{13}$ W/cm$^2$ and high projectile energies.
A hybrid approach is used, where for the interaction of
the incident and scattered  electrons with the laser field we employ the
Gordon-Volkov wave functions, while the interaction of the hydrogen atom
 with the laser field is  treated in second-order perturbation theory.
 Within this formalism, a closed analytical solution is derived for
 the nonlinear  differential cross section,  which is valid for circular as well linear polarizations.
Simple analytical expressions of differential cross section  are derived
in the weak field domain for two-color laser field that is a combination of the fundamental
and its second or third harmonics.
It is shown that the nonlinear differential cross sections  depend
on the dynamical phase of the  scattering process and on
the helicities of the two-color circularly polarized laser field.
A comparison between the two-photon absorption scattering signal for two-color co- and counter-rotating
circularly polarized laser fields is made for even ($2\omega$)
or odd ($3\omega$) harmonics,  and the effect of the intensity ratio of the
 two-color laser field components  is studied.
We analyze the origin of the symmetries in the differential cross sections
and we show that the modification of the photon helicity implies a change
in the symmetries of the scattering signal.

\end{abstract}
\date{\today}
\pacs{34.80.Qb, 34.50.Rk, 32.80.Wr}
\keywords{free-free transitions,elastic scattering,circular polarization,bicircular laser field}
\maketitle

\section{INTRODUCTION}
\label{I}

The study of laser-assisted and laser-induced atomic processes has attracted an increasing theoretical
as well experimental interest in the last 30 years.
In particular, such interest is justified because of the possibility of controlling
the atomic processes by using two-color (or multicolor) fields and manipulating the
 laser parameters such as relative phase and intensities ratio
between the monochromatic components of the fields, polarizations, etc. \cite{ehl2001}.
The physical mechanism for controlling laser-assisted and laser-induced atomic processes using two-color
 fields resides in the different pathways, due to monochromatic components of the field,
 leading to the same final state of the atomic system involving
different numbers of exchanged (absorbed and/or emitted) photons.
Thus, the quantum-mechanical interference occurring among different transition amplitudes,
which contribute to the same final state but through distinct pathways,
offers the possibility of controlling laser-assisted and laser-induced atomic processes.
Different kinds of electronic transitions were investigated, such as
laser-assisted elastic and inelastic electron-atom scattering in  a laser field (free-free transitions),
laser-induced excitation of atoms (bound-bound transitions),
or laser-induced ionization of atoms (bound-free transitions).
It is well known that laser-assisted electron-atom scattering  is a subject of particular interest
 in applied domains such as laser and plasma physics \cite{plasma},  astrophysics \cite{astro},
 or fundamental atomic collision theory \cite{massey}.
Detailed reports on the laser-assisted electron-atom collisions can be found in several review papers
 \cite{mason,ehl1998} and books \cite{bransden,joa2012}.
Recently, the study of electron-atom scattering in the presence of a laser field
has attracted considerable attention \cite{musa,Harak,Kanya,Kanya2,gabi2017},
especially because of the progress of experimental techniques.

A two-color bicircular electromagnetic field represents a superposition of two circularly
polarized (CP) fields which rotate in the same plane, with different photon energies
and the same helicities (corotating CP fields)  or opposite helicities (counter-rotating CP fields).
More than 20 years ago, it was suggested  that CP high-order harmonics  can be generated
 by counter-rotating bicircular fields for a zero-range-potential model atom \cite{Long}
 and  it was experimentally shown that the emission of these harmonics is very efficient compared to
corotating CP and linearly polarized (LP) fields \cite{Eichmann}.
The recent experimental confirmation that the generated high harmonics are
circularly polarized  \cite{Fleischer2014},
allowing the direct generation of CP soft x-ray pulses,
  has generated an increasing interest in studying different
laser-induced processes  by co- and counter-rotating CP laser fields such as
 strong-field ionization \cite{Mancuso15}, nonsequential double ionization \cite{Mancuso16},
or laser-assisted electron-ion recombination \cite{Odzak1,Odzak}.

In the present  paper,  we  study the elastic scattering of fast electrons  by hydrogen atoms
 in their ground state in the presence of  two-color bicircular laser fields.
Obviously, the physical mechanism occurring in laser-assisted processes for two-color fields
 is the interference among different photon pathways leading to the same final state, and by
using CP fields another parameter (the  \textit{dynamical phase}
of the scattering process) plays an important role.
Theoretical studies involving monochromatic CP fields with  the atomic dressing taken
into account in second-order of time-dependent perturbation theory (TDPT) were performed
for laser-assisted electron-hydrogen scattering by Cionga and coworkers \cite{acgabi2,acgabiopt}.
In contrast to these past studies, we investigate here a different regime of a bichromatic CP field
where  the monochromatic components of the  field  rotate in the same plane
and in the same or opposite directions.
To our knowledge, there are no other theoretical studies regarding elastic laser-assisted
 electron-atom scattering processes in a two-color bicircular laser field which include
the atomic dressing in second-order TDPT.
The paper is structured as follows.
In Sec. \ref{II} we present the semiperturbative method used
to  obtain the analytical formulas for the differential cross section (DCS)
for a two-color CP laser field with different polarizations.
Because the scattering process under investigation is a very complex problem, the
theoretical approach poses considerable difficulties and few assumptions are made.
Moderate field intensities below $10^{13}$ W/cm$^2$ and
 fast projectile electrons are considered in order
to safely  neglect the second-order Born approximation
in the scattering potential as well as the exchange scattering \cite{b-j}.
The interaction between the projectile electrons and the laser field
is described by Gordon-Volkov wave functions \cite{volkov}, whereas
the interaction of the hydrogen atom with the laser field  is described
within the second-order TDPT in the field  \cite{vf1}.
The derived  analytical formula for the DCS, which includes the second-order atomic
dressing effects,  is valid for  two-color fields with circular and/or linear polarizations.
In Sec. \ref{III}, we provide  \textit{simple analytic formulas} in a closed form for DCSs
in the laser-assisted elastic scattering  in the  weak field  limit,
for the superposition of the fundamental laser field with its second or  third harmonic,
which exhibit an explicit dependence on the field polarizations.
The numerical results are discussed in Sec. \ref{IV}, where
the DCSs by corotating and counter-rotating  CP  fields
 are compared and analyzed as a function of the scattering and azimuthal
angles of the projectile electron at different intensity ratios  of the monochromatic
components of the bicircular laser field.
Atomic units (a.u.) are used throughout this paper unless otherwise specified.

\section{Semiperturbative theory at moderate laser intensities for a two-color bicircular laser field}
\label{II}

The laser-assisted scattering of electrons by  hydrogen atoms in a two-color laser field is
formally represented as
\begin{eqnarray}
e^-(E_{p},\mathbf{p})
+  {\rm H}(1s) +N_{1i} \, \gamma \,(\omega_1,  \bm{\varepsilon}_1)
+ N_{mi} \, \gamma \,(\omega_m, \bm{\varepsilon}_{m})
\to  \nonumber \\
e^-(E_{p^\prime},{\mathbf{p^\prime}})
+{\rm H}(1s) + N_{1f} \, \gamma \,(\omega_1,   \bm{\varepsilon}_1)
+ N_{mf} \, \gamma \,(\omega_m,  \bm{\varepsilon}_{m} ),
\label{process}
\end{eqnarray}

\noindent
where $ E_{p} \,(E_{p^\prime}) $ and $\mathbf{p} \,(\mathbf{p^\prime}$) denote
the kinetic energy and  the momentum vector of the incident (scattered) projectile electron.
 $\gamma\,(\omega_k, \bm{\varepsilon}_k)$ represents a photon with the energy $\omega_k $
 and the unit polarization vector $\bm{\varepsilon}_{k}$, and
$N_k=N_{ki}-N_{kf}$  denotes the net number of exchanged photons
between the projectile-atom system and each monochromatic component of the two-color laser field,
with $k=1$ and $m$.
The  two-color laser field is treated classically and  is considered as a
superposition of two coplanar CP electric fields,
\begin{equation}
{ \bf E} (t) =
 \frac{i}{2}[
 { E}_{01} \bm{\varepsilon}_1  \exp(-i  \omega_1  t)
+{ E}_{0m} \bm{\varepsilon}_m  \exp(-i  \omega_m  t)]+ \rm{ c.c.},
\label{field}
\end{equation}

\noindent
where ${E}_{0k}$ represents the peak amplitude of the monochromatic components of electric field
and $ \rm{ c.c.}$ denotes the complex conjugate of  the right-hand-side term.
\noindent
For a bicircular field the polarization vector of the first laser beam is defined as
$\bm{\varepsilon}_1 =\bm{\varepsilon}_+= ( \mathbf{e}_{j} +  i \mathbf{e}_{l} ) /\sqrt{2}$,
where $\mathbf{e}_{j}$ and  $\mathbf{e}_{l}$ are the unit vectors along two orthogonal directions,
and the second laser beam has the same polarization
$ \bm{\varepsilon_{m}}  = \bm{\varepsilon}_+$ (the so-called \textit{corotating} polarization),
or   is  circularly polarized in the opposite direction
$ \bm{\varepsilon_{m}}  = \bm{\varepsilon}_{-}=(\textbf{e}_j-i \textbf{e}_l)/\sqrt{2}$
 (the so-called \textit{counter-rotating}  polarization).
Thus, the two-color bicircular electric  field vector  can be easily  calculated as
\begin{equation}
{ \bf E}(t) =
 \mathbf{e}_{j} \left({ E}_{01}\sin \omega_1 t +    { E}_{0m}\sin \omega_m t\right)/\sqrt{2}
-\mathbf{e}_{l} \left({ E}_{01}\cos \omega_1 t +\eta \, { E}_{0m}\cos \omega_m t\right)/\sqrt{2},
\label{cnfield}
\end{equation}
where the helicity takes the values $\eta =+1$ for corotating CP fields
and  $\eta =-1$  for  counter-rotating CP fields.

\subsection{Projectile electron and atomic wave functions}

We consider  moderate laser intensities and fast projectiles,
which imply  that the strength of the laser field is lower than the Coulomb field strength
experienced by an electron in the first Bohr orbit of the hydrogen atom and the
energy of the projectile electron is much higher than the energy of
the bound electron in the first Bohr orbit \cite{ehl1998}.
The interaction between the projectile electron and the two-color  laser field is treated
by a Gordon-Volkov wave function \cite{volkov}, and
the initial and final states of the projectile electron are given by
\begin{equation}
 \chi_{ \mathbf{p}}  ({\mathbf{R}},t)={(2\pi )^{-3/2}}
        \exp \left[ -iE_pt+i{\mathbf{p}}\cdot {\mathbf{R}}
-i  {\mathbf{p}} \cdot \bm{\alpha}_1(t) -i  {\mathbf{p}} \cdot \bm{\alpha}_m(t) \
\right] ,
 \label{fe}
\end{equation}
where $\mathbf{R}$ denotes the position vector  and
$\bm{\alpha}_k(t)$, with $k=1$ and $m$, describes the classical oscillation
motion of the projectile electron in the bicircular electric fields  defined by Eq. (\ref{cnfield}),
\begin{eqnarray}
\bm{\alpha}_{1}(t) &= &\alpha_{01}
\left( \mathbf{e}_{j} \sin \omega_1 t  + \mathbf{e}_{l} \cos \omega_1 t \right)/\sqrt{2},\\
\bm{\alpha}_{m}(t)  &= &\alpha_{0m}
\left( \mathbf{e}_{j} \sin \omega_m t  +\eta \, \mathbf{e}_{l} \cos \omega_m t \right)/\sqrt{2},
\label{quiver}
\end{eqnarray}
 where ${\alpha_{0k}} = \sqrt{{I}_{k}}/ \omega_k^{2}$ is the peak amplitude
and ${I}_{k}={ E}_{0k}^2$ is the peak laser intensity.
In Eq. (\ref{fe}) the terms which are proportional to the ponderomotive
energy $U_{p,k}= {I}_{k}/ 4 \omega_k^{2}$ are neglected
since the calculations  presented in this paper are made at
 moderate field intensities below $10^{13}$ W/cm$^2$.
For example, at a field intensity of $10^{12}$ W/cm$^2$ and a photon energy
 of $1.55$ eV, the  ponderomotive energy
is about  $0.06$ eV  and therefore is negligible in
comparison with the photon and  projectile energies.

At moderate field strengths  the interaction of the hydrogen atom
 with the two-color laser field is considered within the second-order TDPT
and an approximate solution for the wave function in the second-order TDPT
for an electron bound to a Coulomb potential in the presence of an
electric field, Eq. (\ref{field}), is written as
 \begin{equation}
\Psi _{1s}\left( \mathbf{r}, t\right)  =
\exp{(-i { E}_{1s}t)}
\left[
\psi _{1s} (\mathbf{r},t)  + \psi_{1s}^{(1)}(\mathbf{ r},t)
+ \psi_{1s}^{(2)}(\mathbf{ r},t)
\right]\,
,
 \label{fat}
\end{equation}
where  $\bm{r}$ represents the position vector of the bound electron,
 ${E}_{1s}$ is the energy of the ground state, $ \psi _{1s} $
is the unperturbed  wave function of the ground state, and
  $\psi_{1s}^{(1)}$ and  $\psi_{1s}^{(2)}$ represent the first- and second-order
radiative corrections to the atomic wave function.
We employ the following expression of the first-order correction,
\begin{equation}
\psi_{1s}^{(1)}(\mathbf{r},t) =-\sum_{k=1,m}
\frac{\alpha_{0k} \omega_k}{2} \left[
 \bm{\varepsilon}_k\cdot
 \mathbf{w}_{100}(\Omega^{+}_k;\mathbf{r} )  \exp{(-i\omega_k t)} +
   \bm{\varepsilon}_k^* \cdot
 \mathbf{w}_{100}(\Omega^{-}_k;\mathbf{r} )\exp{(i\omega_k t)} \right]
,\label{co1m}
\end{equation}
where  ${\bf{w}}_{100}$ is the  linear-response vector  \cite{vf1},
which depends on the energies $ \Omega_k^{\pm} = {E}_{1s} \pm \omega_k$,
 that is expressed as
\begin{equation}
{\bf{w}}_{100}(\Omega ;\mathbf{r})
 =i (4\pi)^{-1/2}{\cal B}_{101} (\Omega;r) \, \hat{\mathbf{r}}.
\label{wb}
\end{equation}
\noindent
The radial function ${\cal B}_{101}$ is calculated
  using the Coulomb Green's function including both
 bound and continuum eigenstates \cite{vf1} and
 $\hat{\mathbf{r}} = \mathbf{r} /| \mathbf{r}|$.
\noindent
The second-order correction to the atomic wave function $\psi_{1s}^{(2)}$
is written in terms of the quadratic response  tensors \cite{vf2} as,
\begin{equation}
{w}_{jl,100}(\Omega',\Omega ;\mathbf{r})
 =\frac{1}{6 \sqrt{\pi}}
\left\{
\frac{3x_j x_l}{r^2}{\cal B}_{10}^{21} (\Omega',\Omega ;\mathbf{r})
+\delta_{jl} [ {\cal B}_{10}^{01}(\Omega',\Omega ;\mathbf{r})
-{\cal B}_{10}^{21}(\Omega',\Omega ;\mathbf{r}) ]
\right\}
,
\label{wb2}
\end{equation}
 and
\begin{equation}
\widetilde w_{jl,\rm 100} ({E_{1s}}, \Omega ;\mathbf{r})  =
\frac{1}{2}\lim_{\epsilon \to 0}
 \left[ w_{jl,\rm 100} ({ E_{1s}+\epsilon}, \Omega;\mathbf{r})   +
w_{jl,\rm 100} ({E_{1s}-\epsilon}, \Omega;\mathbf{r})  \right],
\label{wt}
\end{equation}
where the radial functions ${\cal B}_{10}^{01}$ and ${\cal B}_{10}^{21} $
are calculated  in Ref. \cite{vf2}, with $j$ and $l =1,2,3$.
The explicit form of the second-order radiative correction  $\psi_{1s}^{(2)}$
for a two-color laser field is given in Appendix \ref{A1}.

\subsection{Scattering matrix}
\label{scm}

To proceed, once we have obtained the atomic and projectile electron wave function
 in the laser field we are able to derive the scattering matrix
for the electron-atom scattering in the static potential
$
V(r,R)=-1/R+ 1/{|\bf{R}-\bf{r}|}.
\label{pot}
$
We assume fast projectile electrons such that the scattering  process can be
well treated within the first-order Born approximation
in the scattering potential $V(r, R)$
and we use a semiperturbative treatment for the
scattering process similar to the one developed by Byron and Joachain \cite{b-j}.
The scattering matrix element is calculated  at  high projectile energies,
$E_{p}>100$ eV, where the exchange effects can be safely neglected, as
\begin{equation}
S_{fi} = -i \int_{-\infty}^{+\infty} dt
\langle \chi_{ \mathbf{p^\prime}}( \mathbf{R},t)
\Psi_{1s}(\mathbf{r},t)  		 |V(r,R) |
 	      {\chi}_{ \mathbf{p}}( \mathbf{R}, t)
\Psi_{1s}(\mathbf{r},t)
			\rangle
\,,
\label{sm}
\end{equation}
\noindent
where $\chi_{\mathbf{p}_i}$  and $\chi_{\mathbf{p^\prime}}$ are  the initial and final
Gordon-Volkov wave functions  of the projectile electron in the two-color laser field
 and $\Psi_{1s} $  represents the  wave function of the bound electron in the  laser field,
 given by Eqs. (\ref{fe}) and (\ref{fat}), respectively.

For commensurate photon energies,  $\omega_m =m \omega_1$,
the energies of the projectile electron satisfy the conservation relation
$ E_{p^\prime}  -   E_{p}  = N \omega_1 $, with $N \omega_1 \equiv N_1 \omega_1 +N_m \omega_m$.
Using the  Jacobi-Anger expansion formula
$
\exp{(i a  \sin \omega  t)} =\sum_{N} J_N(a) \exp{(i N \omega t)},
$
we develop the field dependent part of the Gordon-Volkov wave functions in the scattering matrix,
 Eq. (\ref{sm}), in terms of the phase-dependent generalized Bessel functions,
 $ B_N$, \cite{Watson,varro}, as
\noindent
\begin{equation}
\exp{[-i  \bm{\alpha}_1(t)  \cdot \mathbf{q} -i  \bm{\alpha}_m(t)  \cdot \mathbf{q}]} =
\sum_{N=-\infty}^{+\infty} B_N(R_{1},R_{m};\phi_{1},\phi_{m})
 \exp{(-i N \omega_1 t +i N \phi_{1})},
\label{bgv}
\end{equation}
\noindent
where
\begin{equation}
 B_N(R_{1},R_{m};\phi_{1},\phi_{m})
=\sum_{l=-\infty}^{+\infty}
 J_{N-ml}(R_{1})  J_{l}(R_{m})\exp{[-i  l(m\phi_{1}- \phi_{m})]}.
\label{bn}
\end{equation}
\noindent
 Here $\mathbf{q}$ denotes the momentum transfer vector
of the projectile electron, i.e., $ \mathbf{q}= \mathbf{p} - \mathbf{p^\prime}$,
 and the arguments of the generalized Bessel function are
 $ R_{k}= \alpha_{0k}|\bm{\varepsilon}_k \cdot \mathbf{q} |$
 and  $\phi_{k}$, where the dynamical phase is defined as
$ e^{i \phi_{k}} =  (\bm{\varepsilon}_k\cdot \mathbf{q})
/ |\bm{\varepsilon}_k\cdot \mathbf{q} |$, with $k=1$ and $m$.
Specifically, for a CP field we obtain
$ R_{k}= \alpha_{0k}\sqrt{(\mathbf{e}_j\cdot \mathbf{q})^2+(\mathbf{e}_l\cdot \mathbf{q})^2}/\sqrt{2}$ and $\phi_{k}=\arctan{(\mathbf{e}_l\cdot \mathbf{q})/(\mathbf{e}_j\cdot \mathbf{q})} +s\pi$,
 where $s$ is an integer.
Obviously, a change of helicity of the CP field, i.e., $\bm{\varepsilon} \to \bm{\varepsilon}^*$,
leads to a change of the sign of the dynamical phase, $\phi_{k} \to -\phi_{k}$.
In contrast, for a LP field the arguments of the $B_N$  function are simply  given by
 $ R_k^{LP}=\alpha_{0k} | \mathbf{e}_j\cdot \mathbf{q} |$ and $\phi_k^{LP}=s\pi$.

By substituting Eqs. (\ref{fe}),(\ref{fat}) and  (\ref{bgv}) into
 Eq. (\ref{sm}) we obtain, after integrating over time and
 projectile electron coordinate, the scattering matrix
 $ S_{fi}$ for  elastic electron-hydrogen collisions
in a two-color laser field,

\begin{equation}
S_{fi} =
- 2\pi i \sum_{N=-\infty}^{+\infty}	T_{fi}(N) \,
\delta( E_{p^\prime}  -   E_{p}  - N \omega_1) \,,
\label{smt}
\end{equation}

\noindent
where the Dirac $\delta$ function assures the  energy conservation, and
 $ T_{fi} (N) $ represents the total transition amplitude
for the elastic scattering  process,  which  can be written as the sum of three terms

\begin{equation}
T_{fi}(N) = T^{(0)}(N) + T^{(1)}(N)+ T^{(2)}(N)
.\label{tgen}
\end{equation}
\noindent
Finally, the nonlinear DCS for the scattering of the projectile  in the solid angle
$ \Omega$  with  the energy of the projectile  modified by $ N\omega_1 $, reads
\begin{equation}
\frac{d{\sigma}(N)}{d\Omega} =
 {(2\pi)}^4  \frac{{p^\prime}{(N)}}{p}  {| T^{(0)}(N) + T^{(1)}(N)+ T^{(2)}(N) |}^2
,\label{dcs}
\end{equation}

\noindent
in which the final momentum of the projectile is calculated as
${p^\prime}  (N)= {( p^{2} + 2N \omega_1 ) }^{1/2}$.
The derivation of the transition amplitudes $T^{(i)}(N)$, ($i=0,1,2$)
is briefly described in the next subsections.

\subsubsection{Electronic transition matrix elements}

The first term,  $T^{(0)}$, on the right-hand side of Eq. (\ref{tgen})
represents the elastic  transition amplitude due to projectile electron contribution
in which the atomic dressing is neglected,

\begin{equation}
T^{(0)}(N) =
B_N (R_{1},R_{m};\phi_{1},\phi_{m})
\langle\psi_{1s}|F(\mathbf{q},\mathbf{r})|\psi_{1s}\rangle ,
\label{tn0}
\end{equation}

\noindent
where the  form factor is given by
$ F(\mathbf{q},\mathbf{r}) =
	\left[ \exp{(i  \mathbf{q} \cdot \mathbf{r}) } - 1 \right]  /{(2 \pi^2 q^2)}
\label{ff}$.
By substituting   the partial-wave expansion of the exponential term
$\exp{(i \mathbf{q} \cdot \mathbf{r}) }$ in the  form factor and
 performing the angular integration,
the electronic transition amplitude $T^{(0)}$ reads
\begin{equation}
T^{(0)} (N) =- \frac{1}{(2\pi)^2}  f_{el}^{B_1}(q) \, B_N(R_{1},R_{m};\phi_{1},\phi_{m}),
\label{tne}
\end{equation}
where the term  $ f_{el}^{B_1}(q) ={ 2(q^{2}+8) }/{(q^2+4)^2}$
represents   the first-order Born approximation of the
scattering amplitude for field-free elastic scattering process,
and the laser field dependence  is contained in  the arguments of the
 generalized  Bessel function only.
\noindent
If the atomic dressing is negligible in Eq. (\ref{dcs}),
 the DCS for $N$-photon exchange is approximated as
\begin{equation}
\frac{d{\sigma}(N)}{d\Omega} \simeq
\frac{4p'}{p} \frac{ (q^{2}+8)^2 }{(q^2+4)^4}
\left|B_N(R_{1},R_{m};\phi_{1},\phi_{m}) \right|^2
,
\end{equation}
which is the equivalent of the Bunkin and Fedorov formula \cite{bf},
 for fast electron scattering on hydrogen atoms in a  two-color CP laser field,
 in which the laser-atom interaction  is neglected.

\subsubsection{First-order atomic transition matrix elements}

The second term, $T^{(1)}$, on the right-hand side of the transition amplitude,
 Eq. (\ref{tgen}), represents the first-order atomic transition amplitude and it
occurs due to alteration of the atomic state by the two-color laser field, which is
 described by  the first-order radiative correction $\psi^{(1)}(\mathbf{r},t)$.
Obviously, only one photon is emitted or absorbed
between the two-color  laser field and  the bound electron.
 After some algebra, the atomic transition amplitude $T^{(1)}$ is expressed as
\begin{equation}
T^{(1)}(N) = -\sum_{k=1,m} \frac{\alpha_{0k} \omega_k}{2}
\left[ B_{N-k}\;
{\cal M}_{at}^{(1)} ( \omega_{k} ,\mathbf{q}) e^{-ik\phi_{1}}+
B_{N+k}\;
 {\cal M}_{at}^{(1)} ( -\omega_{k}, \mathbf{q} ) e^{ik\phi_{1}} \right],
\label{tn1}
\end{equation}
\noindent
where  $ {\cal M}_{at}^{(1)}(\omega_k,\mathbf{q})$ denotes a specific
first-order atomic transition matrix element related to one-photon absorption
\begin{equation}
{\cal M}_{at}^{(1)}( \omega_k,\mathbf{q}) = \sum_{j=1}^{3}{\varepsilon}_{kj}
[\langle\psi _{1s}|F(\mathbf{q})|
{{w}}_{j,100}( \Omega_k^{+}) \rangle
+\langle  {w}_{j,100}
( \Omega_k^{-} )|F(\mathbf{q})|\psi _{1s}\rangle],
\label{def1}
\end{equation}
\noindent
whereas the transition matrix element $ {\cal M}_{at}^{(1)} ( -\omega_{k}, \mathbf{q} ) $
 is related to one-photon emission and is obtained from
$ {\cal M}_{at}^{(1)} (\omega_{k}, \mathbf{q} ) $
with the replacements $ \omega_{k} \to -\omega_{k}$ (i.e., $ \Omega_{k}^{\pm} \to \Omega_{k}^{\mp}$)
and $\bm{\varepsilon}_k \to \bm{\varepsilon}_k^*$.
The first term on  the right-hand side of Eq. (\ref{def1})
describes  first the atom interacting  with the laser field followed by
 the   projectile electron-atom interaction, while
in the second term the projectile electron-atom interaction precedes the atom-laser interaction.
For the sake of simplicity, the arguments of the generalized Bessel functions
$B_N (R_{1},R_{m};\phi_{1},\phi_{m}) $ are now dropped off in Eq. (\ref{tn1})  and
throughout this paper.
By using the partial-wave expansion of the exponential term
$\exp{(i  \mathbf{q} \cdot \mathbf{r} )}$
and the definition of ${\mathbf{w}}_{100}$ given by  Eq. (\ref{wb}),
after  performing the angular integration,
 we obtain for the atomic transition matrix element,  Eq. (\ref{def1}),
\begin{equation}
{\cal M}_{at}^{(1)}( \omega_k,\mathbf{q}) =
-\frac{ \bm{\varepsilon}_k  \cdot \hat {\mathbf{q}} }{2{\pi}^{2}q^2 }
\left[{\cal J}_{101}^a(\Omega_k^+,q) -{\cal J}_{101}^a(\Omega_k^-,q)\right]
,
\label{ma}
\end{equation}
\noindent
in which   ${\cal J}_{101}^a $  is  an  atomic radial integral defined as
\begin{equation}
{\cal J}_{101}^a( \Omega_k, q ) =
{\int}_{0}^{+\infty} dr\ r^2 R_{10}(r) j _{1}(qr)
{\cal B}_{101} ( \Omega_k ; r ) ,
\label{Ja1}
\end{equation}
 where $j _{1}(qr)$ represents the spherical Bessel function of first kind
and  $\hat {\mathbf{q}}=\mathbf{q}/|\mathbf{q}|$.
An analytic expression of ${\cal J}_{101}^a $ in terms of hypergeometric functions
 is given by Eq. (36) of Ref.  \cite{acgabi2}.
Finally, the  atomic transition amplitude, $T^{(1)}$, is obtained
by substituting Eq. (\ref{def1}) into Eq. (\ref{tn1})
\begin{eqnarray}
T^{(1)}(N) &=&
\sum_{k=1,m} \frac{\alpha_{0k} \omega_k}{4{\pi}^{2}q^2 }
\left[
  (\bm{\varepsilon}_k  \cdot \hat {\mathbf{q}}) B_{N-k} {\cal J}_{101}(\omega_{k},q)
e^{-i k\phi_1} \right.
\nonumber \\ && \left.
- (\bm{\varepsilon}_k^*\cdot \hat {\mathbf{q}}) B_{N+k} {\cal J}_{101}(\omega_{k},q)
e^{ik\phi_1}\right],
\label{tni}
\end{eqnarray}
where the  radial integral ${\cal J}_{101} $  is calculated   as
 the difference between two atomic radial integrals given by Eq. (\ref{Ja1}),
 ${\cal J}_{101}(\omega_{k},q)=
 {\cal J}_{101}^a(\Omega_k^{+},q)- {\cal J}_{101}^a(\Omega_k^{-},q)$.

In addition, in the low-photon energy limit ($\omega \ll E_n -E_{1s}$),
 the atomic radial integral can be approximated
as  ${\cal J}_{101}(\omega,q) \simeq   \omega \, q \, \alpha_d (q) $,
where $\alpha_d $ represents the dynamic dipole polarizability
due to polarization of the target by the projectile electron \cite{milo,acgabi2,acgabi3}.
The analytical form of the dynamic dipole polarizability  within the
first-order TDPT  \cite{acgabi2} in  low-photon energy limit is calculated as
\begin{equation}
\alpha_d (q,Z) =  \frac{\alpha_s(Z) }{{3[ 1+ q^2/(2Z)^2 ]}^3 }
		  \left[ 1+\frac{2}{1+ q^2/(2Z)^2}\right] \;,
\label{ad}
\end{equation}
\noindent
where $\alpha_s(Z)= 4.5 \, Z^{-4}$  denotes the static dipole
polarizability of a H-like ion in the ground state \cite{bransden}.
 We point out that Eq. (\ref{ad})  describes the target dressing effects at both
small and large scattering angles in contrast to the dynamic dipole
polarizability calculated in Ref. \cite{milo} for the hydrogen atom,
$\alpha_d (q) = \alpha_s /{(1+ q^2/4)}^3$,
which gives accurate results only at small scattering angles.

\subsubsection{Second-order atomic transition matrix elements}

The third term, $T^{(2)}$, on the right-hand side of the transition amplitude,
 Eq. (\ref{tgen}), represents the second-order atomic transition amplitude
and occurs due to  modification of the atomic state by the laser field, described by
 the second-order radiative correction $\psi^{(2)}(\mathbf{r},t)$,
which is given in Appendix \ref{A1}.
In this approach only two photons are exchanged (absorbed, emitted, or absorbed and emitted)
between the two-color CP laser field and  the bound electron.
After some calculation the atomic transition amplitude $T^{(2)}$ reads
\begin{eqnarray}
T^{(2)}(N) &=& \sum_{k=1,m}\frac{ \alpha_{0k}^2 \omega_k ^2 }{4}
    \left\{
 B_{N-2k}         {\cal M}_{at}^{(2)}
            ( \omega_k,  \mathbf{q})  e^{-2ik\phi_{1}}
\right. \nonumber \\
    &&
+
B_{N+2k}        {\cal M}_{at}^{(2)}
            ( - \omega_k, \mathbf{q} ) e^{2ik\phi_{1}}
 \nonumber \\
    &&     \left.
+ B_{N}
 \left[\widetilde {\cal M}_{at}^{(2)} ( { E}_{1s},\omega_k )
               +\widetilde {\cal M}_{at}^{(2)}  ( { E}_{1s},-\omega_k )\right]
\right\}
\nonumber \\
&&+ \frac{\alpha_{01} \alpha_{0m} \omega_1 \omega_m }{4}
    \left[
 B_{N-m-1}
        {\cal N}_{at}^{(2)}
       ( \omega_1, \omega_m, \mathbf{q}) e^{-i(m+1)\phi_{1}}
 \right. \nonumber \\
&& +
B_{N+m+1}
         {\cal N}_{at}^{(2)}
            ( -\omega_1, -\omega_m, \mathbf{q}) e^{i(m+1)\phi_{1}}
\nonumber \\
&&
    +   B_{N-m+1}
            {\cal N}_{at}^{(2)}
              (- \omega_1, \omega_m, \mathbf{q})e^{-i(m-1)\phi_{1}}
 \nonumber \\ && \left.
+  B_{N+m-1}
            {\cal N}_{at}^{(2)}
              ( \omega_1, -\omega_m, \mathbf{q})e^{i(m-1)\phi_{1}}
        \right]
.\label{t2}
\end{eqnarray}
We point out that the explicit or implicit presence of the dynamical phase factors
$ e^{i \phi_1}$ and $ e^{ i \phi_m}$
 in  the electronic and atomic transitions amplitudes,
 Eqs. (\ref{tne}), (\ref{tni})  and (\ref{t2}),  can give  different interference terms in DCS
 for corotating  in comparison to counter-rotating CP field.
\noindent
Specific second-order atomic transition matrix elements related to two-photon exchange
 appear in Eq. (\ref{t2}).
Thus, ${\cal M}_{at}^{(2)} $ is related to absorption of two identical photons of energy
 $\omega_k$ and complex polarization $\bm{\varepsilon}_{k}$,
\begin{eqnarray}
{\cal M}_{at}^{(2)}( \omega_k,\mathbf{q} ) &=&
  \sum_{j,l=1}^{3} \varepsilon_{kj} \varepsilon_{kl}
\left[
\langle w_{j,100}(\Omega_k^{-})|F(\mathbf{q})|   {w}_{l,100}(\Omega_k^{+})\rangle
    \right. \nonumber \\
    && \left. +
\langle \psi_{1s}|F(\mathbf{q})|    w_{jl,100} ( \Omega_k^{\prime \,+}, \Omega_{k}^{+}) \rangle
+   \langle w_{jl, 100}( \Omega_k^{\prime \,-}, \Omega_k^{-})
    |F(\mathbf{q})|\psi_{1s}\rangle \right],
\label{defm}
\end{eqnarray}
with $\Omega_k^{\prime \, \pm}=E_{1s} \pm 2 \omega_k $, $(k =1$ and $m$).
The atomic transition matrix element $ {\cal M}_{at}^{(2)} (-\omega_{k}, \mathbf{q} ) $,
 which is related to emission of two identical photons,  is obtained from Eq. (\ref{defm})
 by replacing  $ \omega_{k} \to -\omega_{k}$ and the components of the polarization vectors
$\varepsilon_{kj(l)} \to \varepsilon_{kj(l)}^*$.
 $\widetilde {\cal M}_{at}^{(2)}( \omega_k,\mathbf{q} )$,
 describes the absorption  followed by emission  of the same photon,
and  is derived using the tensor $\widetilde {w}_{ij,100}(E_{1s},\Omega_k) $,
\begin{eqnarray}
\widetilde {\cal M}_{at}^{(2)}( \omega_k,\mathbf{q} ) &=&
  \sum_{j,l=1}^{3} \varepsilon_{kj}^*\varepsilon_{kl}
\left[
\langle w_{j,100}(\Omega_k^{+})|F(\mathbf{q})|   {w}_{l,100}(\Omega_k^{+})\rangle
    \right. \nonumber \\
    && \left. +
\langle \psi_{1s}|F(\mathbf{q})|\widetilde  w_{jl,100} ( E_{1s}, \Omega_{k}^{+}) \rangle
+   \langle \widetilde w_{jl, 100}(  E_{1s}, \Omega_k^{-}) |F(\mathbf{q})|\psi_{1s}
\rangle \right].
\label{defmt}
\end{eqnarray}
Similarly, $\widetilde {\cal M}_{at}^{(2)}( -\omega_k,\mathbf{q} )$,
 which describes the emission followed by absorption  of the same photon,
is derived from Eq. (\ref{defmt})  by replacing $\omega_{k} \to -\omega_{k}$ and polarizations $\varepsilon_{kj(l)} \to \varepsilon_{kl(j)}^*$.

\noindent
${\cal N}_{at}^{(2)}( \omega_1, \omega_m, \mathbf{q}) $ describes the  absorption
 of two  distinct photons of energies  $\omega_1$ and $\omega_m$,
\begin{eqnarray}
{\cal N}_{at}^{(2)}
     ( \omega_1, \omega_m, \mathbf{q})
&=&( 1+{\cal P}_{1m} )
    \sum_{j,l=1}^{3} \varepsilon_{1j} \varepsilon_{ml}
\left[\langle w_{j,100}(\Omega_1^{-})|F(\mathbf{q})|
    {w}_{l,100}(\Omega_m^{+})\rangle
    \right. \nonumber \\
&& \left.
    +\langle \psi_{1s}|F(\mathbf{q})|
            w_{jl,100} ( { \Omega}^{\prime \,+}, \Omega_{m}^{+}) \rangle +
\langle w_{jl, 100} ( { \Omega}^{\prime \,-}, \Omega_m^{-} )
    |F(\mathbf{q})|\psi_{1s}\rangle \right]
,\label{defn}
\end{eqnarray}
in which  $\Omega^{\prime \, \pm} = E_{1s} \pm ( \omega_1 +\omega_m)$, where
${\cal P}_{1m}$ denotes a permutation operator that interchanges the photon energies
$\omega_{1(m)} \to \omega_{m(1)}$ and polarization vectors
$\bm{\varepsilon}_{1(m)} \to \bm{\varepsilon}_{m(1)}$,
whereas ${\cal N}_{at}^{(2)} (- \omega_1, -\omega_m, \mathbf{q}) $
is connected to emission  of  two  distinct photons  of  energies
 $\omega_1$ and $\omega_m$, and  is calculated from  Eq. (\ref{defn}) by replacing
 $ \omega_{k} \to -\omega_{k}$ and $\bm{\varepsilon}_k \to \bm{\varepsilon}_k^*$ ($k=1,m$).
${\cal N}_{at}^{(2)} (- \omega_1, \omega_m, \mathbf{q}) $
is related to one-photon emission and  one-photon absorption  of  two  distinct photons  of energies
 $\omega_1$ and $\omega_m$, and is as well obtained from  Eq. (\ref{defn}) by replacing
 $ \omega_{1} \to -\omega_{1}$ and $\bm{\varepsilon}_1 \to \bm{\varepsilon}_1^*$,
\begin{eqnarray}
{\cal N}_{at}^{(2)}
   (- \omega_1, \omega_m, \mathbf{q})
&=&  ( 1+{\cal P'}_{1m} )
    \sum_{j,l=1}^{3} \varepsilon_{1j}^* \varepsilon_{ml}
\left[\langle w_{j,100}(\Omega_1^{+})|F(\mathbf{q})|
    {w}_{l,100}(\Omega_m^{+})\rangle
    \right. \nonumber \\
&& \left.
    +\langle \psi_{1s}|F(\mathbf{q})|
            w_{jl,100} ( {\Omega}^{ -}, \Omega_{m}^{+}) \rangle +
\langle w_{jl, 100} ( { \Omega}^{ +}, \Omega_m^{-} )
    |F(\mathbf{q})|\psi_{1s}\rangle \right]
,\label{defnae}
\end{eqnarray}
in which $\Omega^{ \pm} = E_{1s} \pm ( \omega_1 -\omega_m)$ with $m \neq 1$,
where ${\cal P'}_{1m}$ denotes a permutation operator that interchanges the photon energies
$\omega_{1(m)} \to -\omega_{m(1)}$ and polarizations
$\bm\varepsilon_{1}^* \to \bm\varepsilon_{m}$ and
$\bm\varepsilon_{m} \to \bm\varepsilon_{1}^*$.

The  analytic expressions of the second-order atomic transition matrix elements
for two identical photons after performing the angular integration are
given  by Eqs. (A2) and (51)  of Ref.  \cite{acgabi2}
\begin{equation}
{\cal M}_{at}^{(2)}(\omega_k,{\mathbf{q}})=
\frac{(\bm{\varepsilon}_k \cdot \hat {\mathbf{q}})^2}{2 \pi^2 q^2}
 {\cal Q}(\omega_k,{q})
+\frac{ \bm{\varepsilon}_k^2}{2 \pi^2 q^2}
 {\cal P}(\omega_k,{q})
,
\label{m2}
\end{equation}
and
\begin{equation}
\widetilde {\cal M}_{at}^{(2)}(\omega_k,{\mathbf{q}})=
\frac{|\bm{\varepsilon}_k \cdot \hat {\mathbf{q}}|^2}{2 \pi^2 q^2}
 \widetilde{\cal Q}(\omega_k,{q})+
\frac{1}{2 \pi^2 q^2}
 \widetilde{\cal P}(\omega_k,{q})
,
\label{m2t}
\end{equation}
\noindent
in which  $\bm{\varepsilon}_k^2=0$ for a CP field
 and $\bm{\varepsilon}_k^2=1$ for a LP field,  where the specific
 expressions of the polarization-invariant atomic radial integrals
$ {\cal P}$ and ${\cal Q}$ for two-photon absorption or emission are given
  by Eqs. (A3) and (A4) of Ref.  \cite{acgabi2}.
\noindent
A general form of the second-order atomic transition matrix elements for the exchange
 of two  distinct photons after performing the angular integration can be written as
\begin{equation}
{\cal N}_{at}^{(2)}(\omega_j,\omega_l,{\mathbf{q}})=
\frac{(\bm{\varepsilon}_j\cdot \hat {\mathbf{q}})
(\bm{\varepsilon}_l \cdot \hat {\mathbf{q}})}
{2 \pi^2 q^2}  {\cal Q}(\omega_j,\omega_l,{q})
+\frac{\bm{\varepsilon}_j  \cdot \bm{\varepsilon}_l}{2 \pi^2 q^2}
 {\cal P}(\omega_j,\omega_l,{q})
,
\label{n2gen}
\end{equation}
\noindent
with  the replacements $ \omega_{k} \to -\omega_{k}$ and
 $\bm{\varepsilon}_{k} \to \bm{\varepsilon}_{k}^*$ if the photon $k$ is emitted ($k=j,l$).
Note that for corotating bicircular fields  $\bm{\varepsilon}_{+}^2=\bm{\varepsilon}_{-}^2=0$,
while for  counter-rotating bicircular fields
$\bm{\varepsilon}_{+} \cdot \bm{\varepsilon}_{-}=1$.
The specific atomic radial integrals $ {\cal P}$ and ${\cal Q}$ depend on the photon energies
$ \omega_{j}$ and $ \omega_{l}$, and the amplitude of the momentum transfer vector, $q$.
We point out that the general structure of Eq. (\ref{n2gen})
 is also similar for  other processes like
the elastic scattering of photons by hydrogen atoms \cite{gavrila1970},
 two-photon bremsstrahlung  \cite{gavrila},
 elastic x-ray scattering by ground-state atoms \cite{manakov},
two-photon ionization of hydrogen \cite{taieb,fifirig2000}, or
two-photon double ionization  \cite{starace},
with the vector $\hat {\mathbf{q}}$ replaced by specific vectors characteristic to each particular process.

\section{Two-photon scattering processes in the weak field domain}
\label{III}

We aim to provide simple analytical formulas in a closed form for DCSs, that are easy to handle,
which would give a deeper physical insight for electron-hydrogen scattering in a
 two-color bicircular laser field.
However, we note that at high laser intensities
in the domain where the atomic dressing is non-negligible
the calculations of the multiphoton processes ($|N| > 2 $)
require that the laser-atom interaction should be treated at least
to third order in the field.
Whenever the argument of the Bessel function of the first kind is small, i.e.,
$R_{1} \ll 1$ and $R_{m} \ll 1$, which is satisfied at low laser intensities or
at small scattering angles with moderate laser intensities,
 approximate expressions of the generalized Bessel functions,
$B_N(R_1,R_m,\phi_{1}, \phi_{m})$, can be used  according to Appendix \ref{A2}.
For $N > 0 $, by keeping the leading terms in the laser fields in the
transition amplitudes  $T^{(1)}$ and $T^{(2)}$, given by Eqs.  (\ref{tni})  and (\ref{t2}),
and neglecting the terms which are proportional to the higher powers of
the fields,  we get the following approximation formula for the total transition amplitude:
\begin{eqnarray}
T_m(N) & \simeq&
-\frac{1}{4\pi^2}f_{el}^{B_1}(q) B_{N}
+\sum_{k=1,m} \frac{\alpha_{0k} \omega_k}{4{\pi}^{2}q^3 }
  (\bm{\varepsilon}_k  \cdot {\mathbf{q}}) B_{N-k} {\cal J}_{101}(\omega_{k},q)
e^{-i k\phi_1}
 \nonumber \\
&&+
 \sum_{k=1,m}\frac{ \alpha_{0k}^2 \omega_k ^2 }{4}
 B_{N-2k}         {\cal M}_{at}^{(2)}
            ( \omega_k,  \mathbf{q})  e^{-2ik\phi_{1}}
\nonumber \\
&&+
 \frac{\alpha_{01} \alpha_{0m} \omega_1 \omega_m }{4}
   B_{N-m-1}             {\cal N}_{at}^{(2)}
              ( \omega_1, \omega_m, \mathbf{q})e^{-i(m+1)\phi_{1}}
\nonumber \\
&&+
 \frac{\alpha_{01} \alpha_{0m} \omega_1 \omega_m }{4}
   B_{N-m+1}             {\cal N}_{at}^{(2)}
              (- \omega_1, \omega_m, \mathbf{q})e^{-i(m-1)\phi_{1}}
.\label{tep}
\end{eqnarray}
The next step  is to substitute  in Eq. (\ref{tep}) the approximate expressions of the
generalized Bessel functions, $B_N$, which are given in the Appendix \ref{A2},
 and to analyze two simple cases of even and odd harmonics, $m=2$ and $3$.
The total transition amplitude that includes the first- and second-order
dressings in Eq. (\ref{tep}) will be written in the next two subsections in a form that
allows us to analyze the dependence on the dynamical phases $\phi_{1}$ and $\phi_{m}$.

\subsection{Case $m=2$, superposition of a fundamental field and its second harmonic}

Therefore, for two-photon absorption ($N = 2$) when the limit $R_{1(m)} \ll 1$ is taken
for $\omega_2 = 2\omega_1$,  by substituting the appropriate approximation
of the generalized Bessel functions,  Eq. (\ref{bnap}),
 keeping the second-order contributions in the fields,
 and neglecting the higher powers of the fields  the  total  transition
amplitude in the weak field domain is obtained from Eq. (\ref{tep})  as
\begin{equation}
 T_{m=2}(N=2)  \simeq
\alpha_{01}^2 {\cal A}_1(\omega_{1},q) |\bm{\varepsilon}_1 \cdot \mathbf{q}|^2
+
\alpha_{02} {\cal A}_2(\omega_{2},q) |\bm{\varepsilon}_2  \cdot  {\mathbf{q}}|
 e^{-i(2\phi_{1}-\phi_{2}) } \, ,
\label{t2l}
\end{equation}
\noindent
that is the sum of the one- and two-photon transition amplitudes for the processes depicted
in Fig. \ref{fig1}(a), modulated by a phase factor.
The first term on the right-hand side describes two-photon absorption ($\omega_1$),
 whereas the second term describes one-photon absorption ($\omega_2$), as schematically
shown in Fig. 1(a),  with the amplitudes $ {\cal A}_1$ and  $ {\cal A}_2$  defined by
\begin{eqnarray}
 {\cal A}_1(\omega_{1},q) & =&
\frac{1}{8\pi^2}
\left[-\frac{f_{el}^{B_1}}{4}
+ \frac{ \omega_1}{q^3} {\cal J}_{101}(\omega_{1},q)
+\frac{ \omega_1 ^2 }{q^4}   {\cal Q}   ( \omega_1,  \mathbf{q})\right]
,\\
 {\cal A}_2 (\omega_{2},q)& =&
\frac{1}{4\pi^2}
\left[-\frac{f_{el}^{B_1}}{2}
+ \frac{ \omega_2}{q^3 }  {\cal J}_{101}(\omega_{2},q) \right].
\end{eqnarray}

\noindent
In the limit of low-photon energies  ($\omega_{1}$ and $\omega_{2} \ll E_n -E_{1s}$)
the DCS is approximated as
\begin{equation}
\frac{d{\sigma}_{m=2}(N=2)}{d\Omega} \simeq
 \frac{p'}{p}
\left|
\frac{\alpha_{01} ^2}{2}| \bm{\varepsilon}_1 \cdot \mathbf{q} |^2
\left( \alpha_d \frac{\omega_1^2}{q^2} -\frac{f_{el}^{B_1}}{4} \right)
+
\alpha_{02}| \bm{\varepsilon}_2 \cdot \mathbf{q}|
\left( \alpha_d \frac{\omega_2^2}{q^2} -\frac{f_{el}^{B_1}}{2}\right)
 e^{-i(2\phi_{1}- \phi_{2})}
\right|^2,
\label{t2le-soft}
\end{equation}
\noindent
where the dynamic polarizability $\alpha_d$ is given by Eq. (\ref{ad}).
Furthermore, if the atomic dressing is negligible in   Eq. (\ref{t2le-soft})
 the DCS is simply calculated as

\begin{equation}
\frac{d{\sigma}_{m=2}(N=2)}{d\Omega} \simeq
 \frac{p'}{p} \frac{ (q^{2}+8)^2 }{(q^2+4)^4}
\left|
\frac{\alpha_{01} ^2}{4}| \bm{\varepsilon}_1 \cdot \mathbf{q} |^2
+
\alpha_{02}| \bm{\varepsilon}_2 \cdot \mathbf{q}|
 e^{-i(2\phi_{1}- \phi_{2})}
\right|^2,
\label{t2le}
\end{equation}
that is the equivalent of the Bunkin and Fedorov formula \cite{bf}
 for  a two-color CP laser field with different polarizations
and commensurate  photon energies $\omega_1 $ and $ 2\omega_1$.
\noindent
We note that the transition amplitude and the DCSs,
Eqs.  (\ref{t2le-soft}) and (\ref{t2le}),
have a simple dependence on the dynamical phases,
$\phi_{1}$ and $\phi_{2}$, as $e^{-i(2 \phi_1 - \phi_2)}$,
that modulates the quantum interference between the one- and two-photon processes
depicted in Fig. \ref{fig1}(a).

\subsection{Case $m=3$, superposition of a fundamental field and its third harmonic}

\begin{figure}
\centering
\includegraphics[width=3in,angle=0]{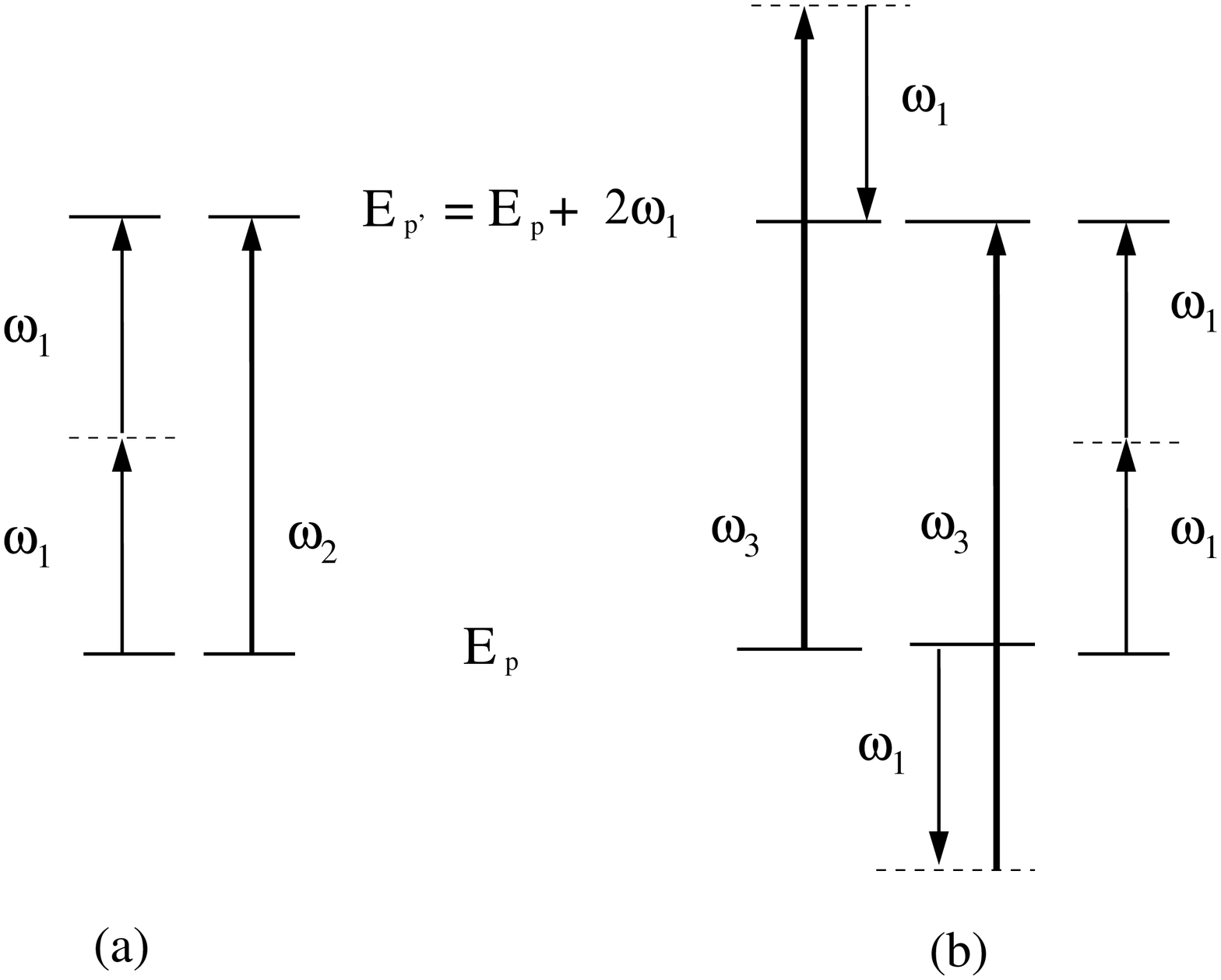}
\caption{ Energy diagrams schematically showing
the channels leading to the final energy of the projectile electron
$E_{p\prime} = E_p + 2\omega_1$.
Channel (a) corresponds to  absorption of two photons of energy $\omega_1$
and one photon of energy $\omega_2=2\omega_1$ of the second harmonic, while channel (b)
 corresponds to  absorption of two photons of energy $\omega_1$
and absorption of the third harmonic $\omega_3=3\omega_1$
and  emission of the  photon $\omega_1$.
}
\label{fig1}
\end{figure}

Similarly, for a two-color laser field that is a superposition of the
 fundamental field and its third harmonic
 $\omega_3 = 3\omega_1$, we get the total transition amplitude for two-photon absorption
in the weak field domain, $R_{1(m)} \ll 1$,
\begin{eqnarray}
 T_3(N=2)  &\simeq&
\alpha_{01}^2 {\cal C}_1(\omega_{1},q) |\bm{\varepsilon}_1 \cdot \mathbf{q}|^2
+
\alpha_{01}\alpha_{03}  \, {\cal C}_2(\omega_1,\omega_3,{q})
|\bm{\varepsilon}_1^*   \cdot  {\mathbf{q}}|
|\bm{\varepsilon}_3  \cdot  {\mathbf{q}}| e^{-i(3\phi_{1}-\phi_{3})}
 \nonumber \\ &&
+\alpha_{01}\alpha_{03}  {\cal C}_3(\omega_1,\omega_3,{q})
(\bm{\varepsilon}_1^*  \cdot \bm{\varepsilon}_3)
 e^{-2i\phi_{1}},
\label{t3l}
\end{eqnarray}
\noindent
that is the sum of the two-photon transition amplitudes for the processes depicted
 in Fig. \ref{fig1}(b),  modulated by  phase factors.
The first term on the right-hand side describes two-photon absorption ($\omega_1$),
 while the rest of the terms describe one-photon absorption ($\omega_3$) and  emission ($\omega_1$),
 as schematically shown in Fig. \ref{fig1}(b),
with  the amplitudes $ {\cal C}_1$, $ {\cal C}_2$,  and  $ {\cal C}_3$  defined by
\begin{eqnarray}
{\cal C}_1(\omega_{1},q)  & =&
\frac{1}{8\pi^2} \left[
-\frac{f_{el}^{B_1}}{4} +\frac{\omega_1}{q^3}{\cal J}_{101}(\omega_{1},q)\right],
\\
{\cal C}_2(\omega_1,\omega_3,{q})  & =&
\frac{1}{8\pi^2} \left[
\frac{f_{el}^{B_1}}{2} -\frac{\omega_3}{q^3}{\cal J}_{101}(\omega_{3},q)
+\frac{ \omega_1 \omega_3 }{ q^4}{\cal Q}(-\omega_1,\omega_3,{q})
\right],
\\
{\cal C}_3(\omega_1,\omega_3,{q}) & =&\frac{ \omega_1 \omega_3 }{8 \pi^2 q^2}
{\cal P}(-\omega_1,\omega_3,{q}).
\end{eqnarray}
\noindent

\noindent
In the limit of low-photon energies ($\omega_{1}$ and $\omega_{3} \ll E_n -E_{1s}$)
the DCS is approximated as
\begin{eqnarray}
\frac{d{\sigma}_{m=3}(N=2)}{d\Omega} & \simeq&
 \frac{p'}{4p}
\left|
 \alpha_{01}\alpha_{03}
|\bm{\varepsilon}_1 \cdot \mathbf{q} ||\bm{\varepsilon}_3 \cdot \mathbf{q}|
\left(  \alpha_d \frac{\omega_3^2}{q^2} - \frac{f_{el}^{B_1}}{2}\right)
e^{-i(3\phi_{1}- \phi_{3})} \right. \nonumber \\&& \left.
-
\alpha_{01}^2| \bm{\varepsilon}_1 \cdot \mathbf{q} |^2
\left(  \alpha_d \frac{\omega_1^2}{q^2}  - \frac{f_{el}^{B_1}}{4}\right)
\right|^2.
\label{t3le-soft}
\end{eqnarray}

\noindent
 Moreover, if the atomic dressing is negligible in Eqs. (\ref{t3le-soft})
the DCS is simply given by
\begin{equation}
\frac{d{\sigma}_{m=3}(N=2)}{d\Omega} \simeq
 \frac{p'}{p} \frac{ (q^{2}+8)^2 }{(q^2+4)^4}
\frac{\alpha_{01}^2| \bm{\varepsilon}_1 \cdot \mathbf{q} |^2}{4}
\left|
 \alpha_{03}|   \bm{\varepsilon}_3 \cdot \mathbf{q}|
e^{-i(3\phi_{1}- \phi_{3})}
-
\frac{\alpha_{01}}{2}| \bm{\varepsilon}_1 \cdot \mathbf{q} |
\right|^2,
\label{t3le}
\end{equation}
that is the equivalent of the Bunkin and Fedorov formula \cite{bf}
 for  a two-color CP laser field with different polarizations
and commensurate  photon energies $\omega_1 $ and $ 3\omega_1$.
Similarly to  the case  $\omega_2= 2\omega_1$,
the  DCS  has a simple dependence on the
 dynamical phases, $\phi_{1}$ and $\phi_{3}$,  as
 $e^{-i(3 \phi_1 - \phi_3)}$, that modulates the quantum interference among
the different two-photon processes depicted  in Fig. \ref{fig1}(b).

\section{NUMERICAL EXAMPLES AND DISCUSSION}
\label{IV}

\begin{figure}
\centering
\includegraphics[width=3.5in,angle=0]{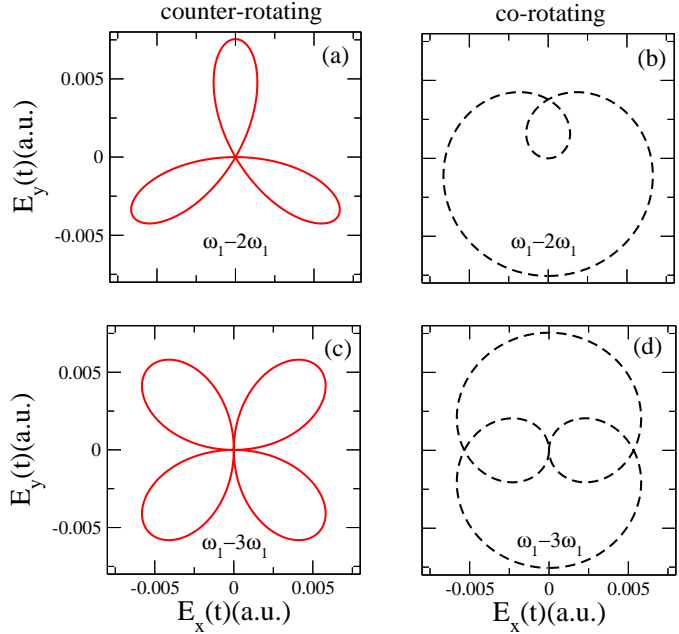}
\caption{(Color online) Parametric plots showing the Cartesian components of the electric field vector
 in the   $(x,y)$-polarization plane,
 $E_x(t)$ and $E_y(t)$, given by Eq. (\ref{cnfield}),  plotted for $0 \leq t \leq T_1$
 at equal laser field intensities  $I_1=I_m=10^{12}$ W/cm$^2$, ($m=2$ and $3$),
for  a  two-color left-handed CP field in the right column with
$\bm{\varepsilon}_{1}= \bm{\varepsilon}_{+}=(\mathbf{e}_x+i\mathbf{e}_y)/\sqrt{2}$ and
$\bm{\varepsilon_{m}}  = \bm{\varepsilon}_{+}$,
and a left- and right-handed CP field in the  left column with
$\bm{\varepsilon_{m}}  = \bm{\varepsilon}_{-}$.
The fundamental photon energy is $\omega_1= 1.55 $ eV
while the energy of the harmonic photon  is $\omega_2=2\omega_1$
 in panels (a) and (b), and $\omega_3=3\omega_1$ in  panels (c) and (d).
The two-color bicircular electric field satisfies a $T_1/(m -1)$ and $T_1/(m+1)$
rotational symmetry for co- and counter-rotating polarizations, respectively.
}
\label{fig2}
\end{figure}

In this section, we present representative results for the
 the scattering process described by Eq. (\ref{process}), in which two photons
are absorbed by the $e^- + {\rm H}(1s)$ colliding system.
We focus our discussion on two particular field polarizations in which the two-color
laser beams are CP in the $(x,y)$ plane
with one laser beam propagating in the $z$-axis direction,
$ \bm{\varepsilon}_{1}  =\bm{\varepsilon}_{+}  =(\mathbf{e}_x+i \mathbf{e}_y)/\sqrt{2}$,
while the other laser beam  (a) has the same (left-handed) circular polarization
$ \bm{\varepsilon_{m}}  = \bm{\varepsilon}_{+}$,
 i.e., \textit{corotating polarization} case (LHCP-LHCP),
or  (b) is (right-handed) CP in the opposite direction,
$ \bm{\varepsilon_{m}}  = \bm{\varepsilon}_{-}=(\mathbf{e}_x-i \mathbf{e}_y)/\sqrt{2}$,
i.e.,   \textit{counter-rotating polarization} case (LHCP-RHCP).
In this polarization geometry for the simple case of equal intensities
of the monochromatic field components,  ${ E}_{01}={ E}_{0m}$,
 the bicircular electric field can be written as
\begin{equation}
{\bf E}_{+} (t) ={ E}_{01} \sqrt{2} \,
(\mathbf{e}_x \sin \omega_+t -\mathbf{e}_y \cos \omega_+t)
\cos \omega_-t,
\label{co}
\end{equation}
for corotating circular polarizations ($ \bm{\varepsilon_{m}}  = \bm{\varepsilon}_{+} $), and
\begin{equation}
{\bf E}_{-} (t) ={ E}_{01} \sqrt{2}\,
(\mathbf{e}_x \cos \omega_-t -\mathbf{e}_y \sin \omega_-t)  \sin \omega_+t
,\label{cn}
\end{equation}
for counter-rotating circular polarizations ($ \bm{\varepsilon_{m}}  = \bm{\varepsilon}_{-} $),
 in which the energies are defined by
$\omega_{\pm}=\eta_m \, \omega_1/2 $,
 where $\eta_m= m -1$ for corotating CP fields and $\eta_m= m +1$
 for counter-rotating CP fields.
\noindent
In Fig. \ref{fig2} is plotted the temporal dependence of the electric field vectors,
given by   Eqs. (\ref{co})  and (\ref{cn}),
in the polarization plane for two-color CP laser fields of equal intensities
with identical polarizations
$ \bm{\varepsilon_{m}}  = \bm{\varepsilon}_{+}$ in the right column, and
$ \bm{\varepsilon_{m}}  = \bm{\varepsilon}_{-}$ in the left column.
The parameters we employ for Fig. \ref{fig2} are   $I_1=I_m=10^{12}$ W/cm$^2$,
a fundamental photon energy $\omega_1=1.55$ eV,  with
$\omega_2=2\omega_1$ in Figs. \ref{fig2}(a) and \ref{fig2}(b), and
$\omega_3=3\omega_1$ in Figs. \ref{fig2}(c) and \ref{fig2}(d).
The electric field vectors are invariant with respect to  translation in time by
an integer multiple of $T_1 /\eta_m$, where $T_1 = 2\pi/\omega_1$ is the fundamental
field optical period, and with respect to rotation in the polarization plane
 by an angle $\alpha_m=2\pi/\eta_m$ around the $z$ axis, such that
\begin{equation}
{\bf E}_{\pm}\left(t + \frac{T_1 }{\eta_m} \right) =
{\bm R }\left(\frac{2\pi }{\eta_m} \right)\, {\bf E}_{\pm}(t),
\end{equation}
where ${\bm R }(\alpha_m)$ is a $2 \times 2$ rotation matrix with angle $\alpha_m$ around the $z$ axis.
 The upper sign (+) corresponds to corotating CP fields,
while the lower sign (-) corresponds to counter-rotating CP fields.
For counter-rotating bicircular field  with $\omega_2=2\omega_1$,
this temporal symmetry means that
$  {\bf E}_-(t + T_1 /3) = {\bm R } (2\pi /3)  \, {\bf E}_-(t)$,
 i.e.,  the translation in time is  one-third of the optical cycle
and the rotation angle in the polarization plane is $2\pi/3$,
which implies a threefold symmetry of the electric field in Fig. \ref{fig2}(a),
while for corotating bicircular field the translation in time is $T_1$
and the rotation angle is $2\pi$ in Fig. \ref{fig2}(b), such that
$  {\bf E}_+(t + T_1 ) = {\bm R } (2\pi )  \, {\bf E}_+(t)$.

\begin{figure}
\centering
\includegraphics[width=2.5in,angle=0]{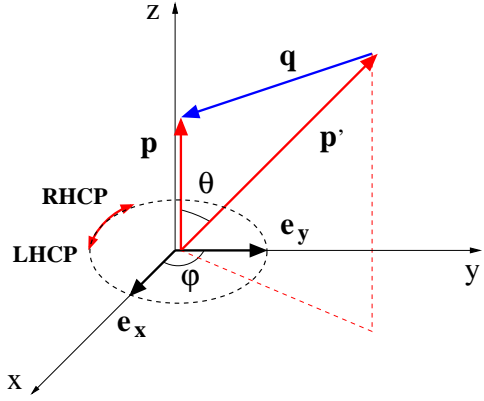}
\caption{
(Color online) Specific scattering geometry  assumed for the numerical
 calculations of laser-assisted $e^-$-H($1s$)  scattering in a bicircular laser field.
We consider the scattering geometry with $\mathbf{p} \parallel \mathbf{e}_z $,
where $\mathbf{p}$ and $\mathbf{p}^\prime$ are the  momentum vectors of the
incident and scattered  electron, $\theta$ is the angle between them,
$\varphi$ is the azimuthal angle, and $\mathbf{q} $ is the momentum transfer vector.
The laser field propagates along the $z$ axis and
is circularly polarized in the $(x,y)$ plane, with the polarization vectors
$\bm{\varepsilon}_{+}=(\mathbf{e}_x+i\mathbf{e}_y)/\sqrt{2}$ (LHCP) and
$\bm{\varepsilon}_{-}=(\mathbf{e}_x - i\mathbf{e}_y)/\sqrt{2}$ (RHCP).
}
\label{fig3}
\end{figure}

Next, we consider the scattering geometry depicted in Fig. \ref{fig3}
in which the momentum vector of the incident electron  $\mathbf{p}$ is parallel
to the $z$ axis.
The momentum transfer vector is given by
$ \mathbf{q}=(-p^\prime \sin \theta \cos\varphi, -p^\prime \sin \theta \sin\varphi,
p -p^\prime \cos \theta )$,
 with the amplitude
$q=\sqrt{ p^2+ {p^\prime}^2 -2 p {p^\prime} \cos \theta}$,
 where $\theta$ is the the scattering angle between the momentum vectors
of the incident and scattered electron, ${\bf p}$ and ${\bf p^\prime}$,
and $\varphi$ is the azimuthal angle of the scattered electron.
In this scattering geometry the scalar product in the argument
of the generalized Bessel functions, $R_k$, can be simply expressed as
$ \bm{\varepsilon}_{\pm}\cdot \mathbf{q}
= - {p^\prime} \sin \theta   /\sqrt{2}\, e^{\pm i \varphi}$.
The  evaluation of the dynamical phases of the  CP laser fields gives
$ \phi_{\pm}   =   \pi \pm  \varphi$ with
$ e^{ i \phi_{\pm}}   =   - e^{\pm i \varphi}$, and
we observe that a change of the field helicity  implies
 a change in the sign of the azimuthal angle $\varphi$.
After some simple algebra,  by replacing the exponential term
 in the  weak field limit of the total transition amplitudes, Eqs. (\ref{t2l}) and (\ref{t3l}),
as $ e^{ -i (m\phi_1- \phi_m)}  =  (-1)^{m+1} e^{- i \eta_m \varphi}$,
  we obtain
\begin{equation}
 T_{m=2}(N=2)  \simeq
 \alpha_{01}^2 {\cal A}_1(\omega_{1},q) |\bm{\varepsilon}_1 \cdot \mathbf{q}|^2
-\alpha_{02}   {\cal A}_2(\omega_{2},q) |\bm{\varepsilon}_1 \cdot \mathbf{q}|
e^{-i \eta_2 \varphi}
,
\label{t2lp}
\end{equation}
\noindent
for $\omega_2=2\omega_1$,  with the parameter $\eta_2= 1$ for two-color left-handed-CP  fields
(equal helicities)
or  $\eta_2= 3$ for two-color left- and right-handed CP fields (opposite helicities), whereas for  $\omega_3=3\omega_1 $
\begin{eqnarray}
 T_{m=3}(N=2)  &\simeq&
\alpha_{01}^2 |\bm{\varepsilon}_1 \cdot \mathbf{q}|^2 \left[ {\cal C}_1(\omega_{1},q)
+
\frac{\alpha_{03}}{\alpha_{01}}  {\cal C}_2(\omega_1,\omega_3,{q}) e^{-i \eta_3 \varphi}
 \right] \nonumber \\ &&+
 \delta_{\eta_3 2} \, \alpha_{01}\alpha_{03}  \, {\cal C}_3(\omega_1,\omega_3,{q})
e^{-2i\varphi}
 ,
\label{t3lp}
\end{eqnarray}
\noindent
with the parameter $\eta_3= 2$ for two-color  left-handed CP fields (equal helicities) or
 $\eta_3= 4$ for two-color left- and right-handed CP fields (opposite helicities), respectively.
Therefore, the transition amplitudes in the weak field domain, given by
Eqs. (\ref{t2lp})  and (\ref{t3lp}), as well
 the corresponding DCSs,
 depend on  the azimuthal angle of the scattered projectile
 as $e^{-i(m -  1)\varphi}$ and are  invariant with respect to the transformations
 $ \varphi \to \varphi + 2\pi/{(m -  1)}$, for corotating polarizations.
In contrast, the transition amplitudes and DCSs for counter-rotating CP fields
depend on  azimuthal angle as $e^{-i(m + 1)\varphi}$
and are  invariant with respect to the transformations
  $ \varphi \to \varphi + 2\pi/{(m +  1)}$.

\noindent
Furthermore, if the atomic dressing  is negligible in Eqs. (\ref{t2lp}) and (\ref{t3lp}),
 the following simple analytic results are obtained for DCSs:
\begin{equation}
\frac{d{\sigma}_{m=2}(N=2)}{d\Omega} \simeq
 \frac{p'}{p} \frac{ (q^{2}+8)^2 }{(q^2+4)^4}
 \frac{\alpha_{01}^2 | \bm{\varepsilon}_1 \cdot \mathbf{q} |^2}{16}
\left|     \alpha_{01} | \bm{\varepsilon}_1 \cdot \mathbf{q} |
- 4  \frac{\alpha_{02}}{\alpha_{01}}  e^{-i \eta_2 \varphi}  \right|^2
,
\label{tne2}
\end{equation}

\noindent
and
\begin{equation}
\frac{d{\sigma}_{m=3}(N=2)}{d\Omega} \simeq
 \frac{p'}{p} \frac{ (q^{2}+8)^2 }{(q^2+4)^4}
\frac{\alpha_{01}^4| \bm{\varepsilon}_1 \cdot \mathbf{q} |^4}{16}
\left|1 + 2  \frac{\alpha_{03}}{\alpha_{01}} e^{-i \eta_3 \varphi}   \right|^2
.
\label{tne3}
\end{equation}
\noindent
We stress that the analytical formulas obtained  for co- and counter-rotating
 circular polarizations  in the weak-field domain,  Eqs. (\ref{t2lp}) and (\ref{t3lp}),
or for negligible atomic dressing,   Eqs. (\ref{tne2}) and (\ref{tne3}),
indicate that DCSs are  invariant with respect to rotation
of the projectile momentum in the polarization plane
  by azimuthal angles $ 2\pi/{(m -  1)}$ and $ 2\pi/{(m +  1)}$ about the $z$ axis, respectively.
It is obvious that  the DCS  by the two-color bicircular laser field with $m=2$
in Eq. (\ref{tne2}) presents the maximal interference between
the first- and second-order transition amplitudes
  at the following optimal ratio of the two-color laser intensities
$I_2 / I_1 \simeq I_1 | \bm{\varepsilon}_1 \cdot \mathbf{q} |^2/ \omega_1^4$,
 which is very sensitive to the  intensity and photon energy of the fundamental field,
and the scattering geometry.
In contrast, for $m=3$ the maximal  interference
in DCS, Eq. (\ref{tne3}),  between the second-order transition amplitudes occurs
 at the optimal ratio of the two-color laser intensities $I_3 /I_1 \simeq 20.2$,
that shows the interference between the different two-photon processes
is more efficient compared to  $m=2$, being independent on the scattering geometry.

\begin{figure}
\centering
\includegraphics[width=4.0in,angle=0]{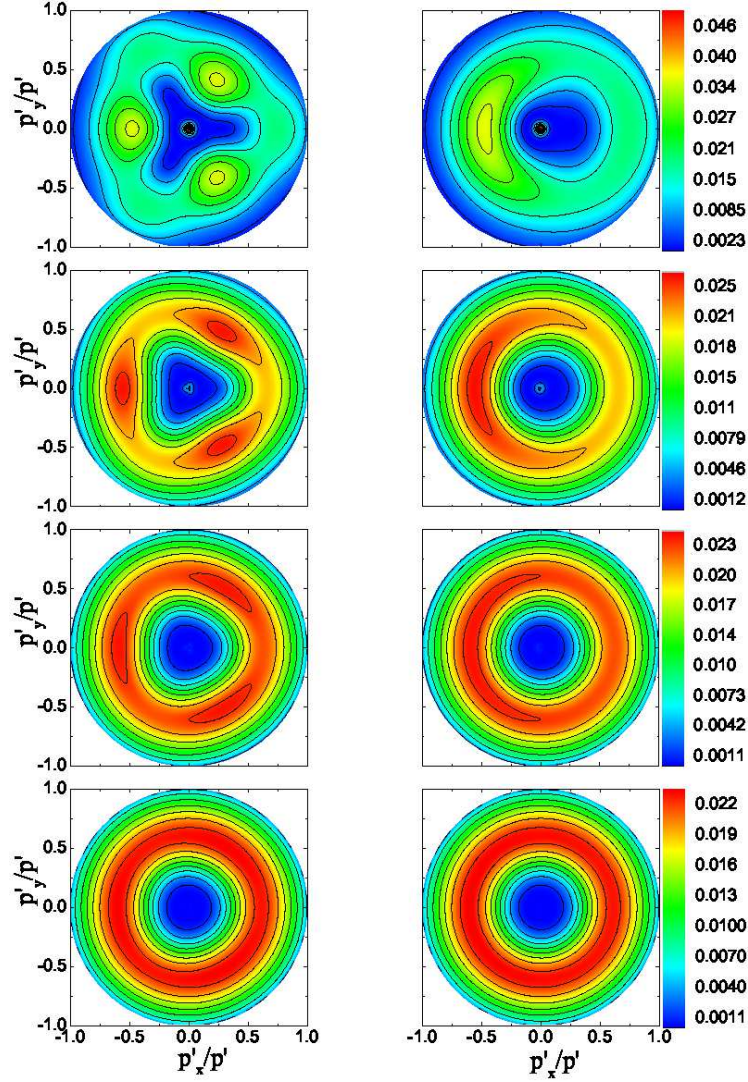}
\caption{(Color online)
Contour plots representing  DCSs ($N=2$),
given by Eq. (\ref{dcs}),  for   a  two-color left-handed CP field in the  right column
and a two-color left- and right-handed CP field in the left column,
  as a function of the normalized  Cartesian  components of the projectile momentum vector
 in the polarization plane, $p_{x}^\prime/{p^\prime} $ and $p_{y}^\prime/{p^\prime} $.
The projectile electron energy is $ E_{p}=100$ eV, $\textbf{p} \parallel \mathbf{e}_{z}$,
and the laser field intensities are
$I_2=I_1,10^{-1}I_1,10^{-2}I_1,$ and $10^{-3}I_1$ from top to bottom,
 with $I_1=10^{12}$ W/cm$^2$.
 The photon energies are $\omega_1= 1.55 $ eV and $\omega_2=2\omega_1$.
The magnitudes of the DCSs in a.u.  are indicated by the color scales in each row.
}
\label{fig4}
\end{figure}

First, we have checked that the two-photon DCSs for the elastic scattering
of fast electrons  by hydrogen atoms in their ground state are in very good agreement
 with the earlier  analytical  data  obtained for the particular case of
a monochromatic CP laser field \cite{acgabi2} and a bichromatic LP laser field \cite{acgabi99}.
Next, we apply the analytic formulas derived in Sec. \ref{II} to evaluate the DCSs
  for  elastic  electron scattering by a hydrogen atom
 in its ground state in the presence of  two-color co- and counter-rotating CP laser fields.
Since our formulas are derived up to  second order in the field for the atomic dressing,
  we analyze two-photon absorption DCSs ($N = 2$) at moderate field intensities.
We choose  high energies of the projectile electron and low-photon energies, such that
neither  the projectile electron nor  the photon can separately excite an upper atomic state.
To start with a simple case we present in Fig. \ref{fig4} our numerical results
for an initial scattering energy $ E_{p}=100$ eV,
 photon energies that correspond to the Ti:sapphire laser $\omega_1 =1.55$ eV
and its second harmonic $\omega_2= 2\omega_1 $, and a fundamental laser
 intensity $I_1=10^{12}$ W/cm$^2$.
These laser  parameters correspond to a quiver motion amplitude
$\alpha_{01} \simeq 1.65$ a.u. and an argument of the Bessel function
$R_1 \simeq  1.65 |\bm{\varepsilon}_1\cdot {\mathbf{q}}|$.
The  intensity of the second-harmonic laser is given by $I_2= f I_1$,
 with the laser intensity ratios
$f = 1,10^{-1},10^{-2},$ and $10^{-3}$ from top to bottom,
which results in a quiver motion amplitude $\alpha_{02} = \alpha_{01} \sqrt{f}/4$
and an argument of the Bessel function $R_2 = R_1 \sqrt{f}/4$.
The three-dimensional electron projectile DCSs, projected in the polarization plane
as a function of the normalized projectile momentum,
$p_{x}^\prime/{p^\prime} $ and $p_{y}^\prime/{p^\prime} $, are plotted
 for two-color left-handed CP fields (corotating)  in the right column
 and for left- and right-handed CP fields (counter-rotating) in the left column.

\noindent
In the weak field domain,
 as long as $\omega_1 < |E_{1s}|/2$ and $\omega_2=2\omega_1$, the one- and two-photon
atomic transition matrix elements
$ {\cal M}_{at}^{(1)}, {\cal M}_{at}^{(2)}$, and $ {\cal N}_{at}^{(2)}$ are real
and  the DCS   can be formally expressed
from Eqs. (\ref{dcs}) and (\ref{t2lp}) as a function of the scattering $ \theta $
and azimuthal $\varphi$ angles,
\begin{equation}
\frac{d{\sigma}_{m=2}(N=2)}{d\Omega}( \theta,\varphi ) \simeq
a_1 \sin^4\theta  +a_2\sin^2\theta - a_3 \sin^3\theta \cos (\eta_2\varphi)
,\label{dcs2p}
\end{equation}
where $a_1=4\pi^4 \alpha_{01}^4  \, {\cal A}_1^2 \, p^{\prime 5}/p $,
$a_2=8\pi^4 \alpha_{02}^2   \, {\cal A}_2^2 \, p^{\prime 3}/p $,
and $a_3=(2\pi)^4 \alpha_{01}^2 \alpha_{02}  \, {\cal A}_1 {\cal A}_2  \, p^{\prime 4}/( p \, \sqrt 2 ) $.
The last term in the right-side hand $a_3 \sin^3\theta \cos (\eta_2\varphi)$,
with $\eta_2=1$ for equal helicities and $\eta_2=3$ for opposite helicities,
 describes the coherent interference between
the first- and the second-order transition amplitudes in Eq. (\ref{t2lp}),
 as schematically shown by the one- and two-photon pathways in Fig. \ref{fig1}(a).
We observe that the DCS, Eq. (\ref{dcs2p}), is invariant to the following transformations:
(i) $\varphi \to \varphi +2\pi/\eta_2$, that is equivalent to a rotation
 in the  azimuthal plane by an  $2\pi/\eta_2$   angle around the $z$ axis
and  (ii) $ \pi-\varphi \to \pi+\varphi $, that is equivalent to a reflection
 with respect to the $(x,z)$ plane.

\begin{figure}
\centering
\includegraphics[width=4.5in,angle=0]{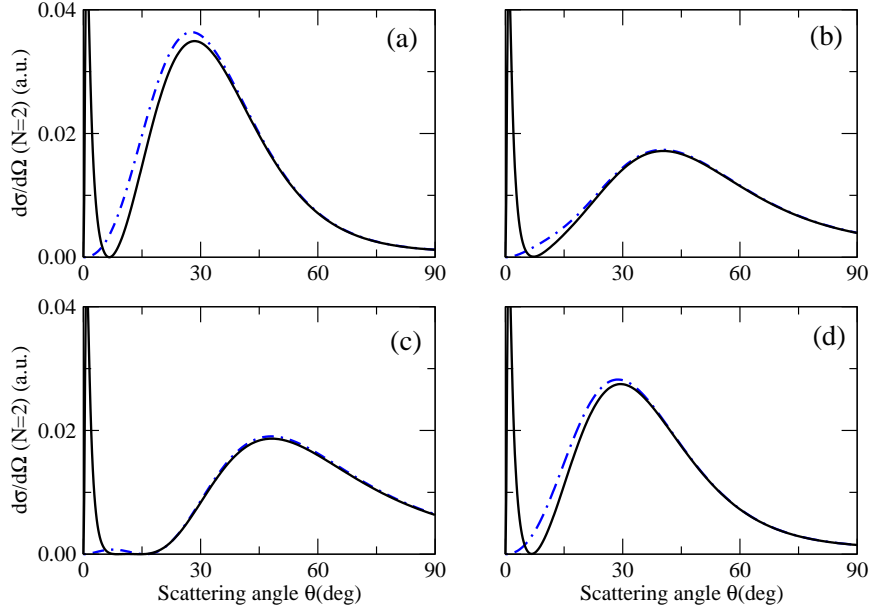}
\caption{(Color online)
Differential cross sections for  $N=2$ by a two-color left- and right-handed-CP
 laser field in panels (a) and (c)
and a two-color left-handed-CP  laser field  in panels (b) and (d)
as a function of the scattering angle  $\theta$, at the azimuthal angle
$\varphi=60^\circ$ in panels (a) and (b), and $\varphi=240^\circ$ in  panels (c) and (d)
for $I_1=I_2=10^{12}$ W/cm$^2$.
The dot-dashed lines represent the projectile electron contribution,
calculated as $(2\pi)^4 (p'/p )|T^{(0)}(N=2)|^2$.
 The rest of the parameters are the same as in  Fig. {\ref{fig4}}.
}
\label{fig5}
\end{figure}

\noindent
For a better understanding of the numerical results presented in Fig. \ref{fig4}  we show
the DCSs as a function of the  scattering angle $\theta$ for $I_2=I_1=10^{12}$ W/cm$^2$,
at the  azimuthal angles $\varphi = 60^{\circ} $
in Figs. \ref{fig5}(a) and \ref{fig5}(b) and $\varphi =240^{\circ} $
 in  Figs. \ref{fig5}(c) and \ref{fig5}(d),
  while the rest of the parameters are the same as in  Fig. \ref{fig4}.
For the employed laser parameters, the laser-atom interaction is quite strong at very small
 scattering angles, $\theta<7^\circ $, only.
Excepting the forward scattering angles, the projectile electron is scattered with
a high probability at $\theta \simeq 40^\circ $ and $\theta \simeq 46^\circ $
in Figs. \ref{fig5}(a) and \ref{fig5}(d)
and at $\theta \simeq 29^\circ $ in Figs. \ref{fig5}(b) and \ref{fig5}(c).
At larger  scattering angles, the projectile electron contribution is
dominant due to nuclear scattering and  determines the angular distribution
of the total DCS.
Very recently, there have been reported new experimental observations  of
the very sharp peak profile at forward scattering angles for $e^-$-Xe scattering
 in LP fields and a few attempts to explain it based on the Zon's model \cite{Kanya2},
which involves the static  polarizability.

\begin{figure}
\centering
\includegraphics[width=5in,angle=0]{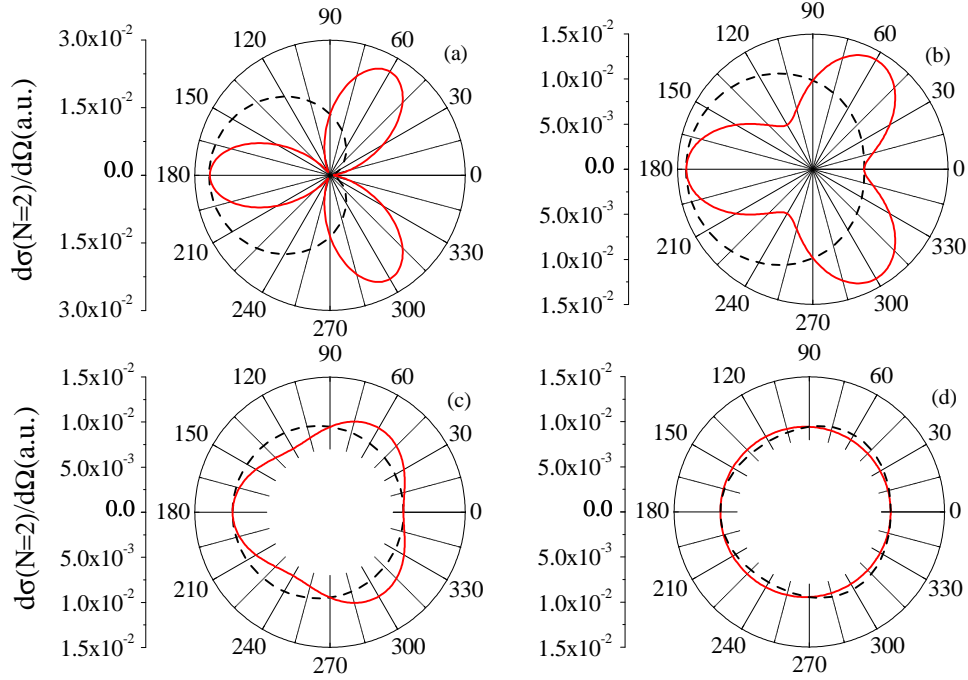}
\caption{(Color online)
Differential cross sections for  $N=2$ by a two-color left- and right-handed-CP (full line) laser field
and a two-color left-handed-CP (dashed line) laser field
as a function of the  azimuthal angle $\varphi$, at  the scattering angle $\theta=20^\circ$,
 for $I_2=I_1$ in panel (a), $10^{-1}I_1$ in panel (b), $10^{-2}I_1$ in panel (c),
 and $10^{-3}I_1$ in panel (d).
 The rest of the parameters are the same as in  Fig. {\ref{fig4}}.
The DCSs by the two-color corotating CP fields are invariant to rotation by an $2\pi$ angle while
by the counter-rotating CP fields are invariant to rotation by an $2\pi/3$ angle.
}
\label{fig6}
\end{figure}

\noindent
In order clarify the origin of the symmetry patterns in Fig. \ref{fig4} for $\omega_2=2\omega_1 $,
 we show DCSs in Fig. \ref{fig6}  for counter-rotating (solid lines) and corotating (dashed lines)
CP fields  as a function of the azimuthal angle $\varphi$,
 at the scattering angle $\theta = 20^\circ $, $I_1=10^{12}$ W/cm$^2$,
 and the harmonic field intensity
$I_2= f I_1$,  with the laser intensity ratios
$f =1$ in Fig. \ref{fig6}(a), $10^{-1}$ in Fig. \ref{fig6}(b),
 $10^{-2}$ in Fig. \ref{fig6}(c), and $10^{-3}$ in Fig. \ref{fig6}(d).
For counter-rotating CP fields we found that the projectile electron is scattered
with a  high probability in the directions of the azimuthal angles
$ \varphi=60^\circ, 180^\circ$, and  $300^\circ$.
 DCS has a specific \textquotedblleft three-leaf clover\textquotedblright pattern, as
described in the weak field domain by the $ \cos 3\varphi $ term  in Eq. (\ref{dcs2p}),
and  is invariant  to rotation around the $z$ axis by an azimuthal angle $ \varphi=2\pi/3$.
In contrast, for the corotating CP fields DCS, has a  pattern described  by the $ \cos \varphi$
 term in Eq. (\ref{dcs2p}) and the invariant rotation angle is  $ \varphi=2\pi$.
For both co- and counter-rotating CP fields, the DCSs are symmetric with respect
to reflection in the $(x,z)$ plane, such that
${d{\sigma}_{m=2}}( \theta, \pi-\varphi )/{d\Omega}
 ={d{\sigma}_{m=2}}( \theta, \pi+\varphi )/{d\Omega}$.

\begin{figure}
\centering
\includegraphics[width=4.0in,angle=0]{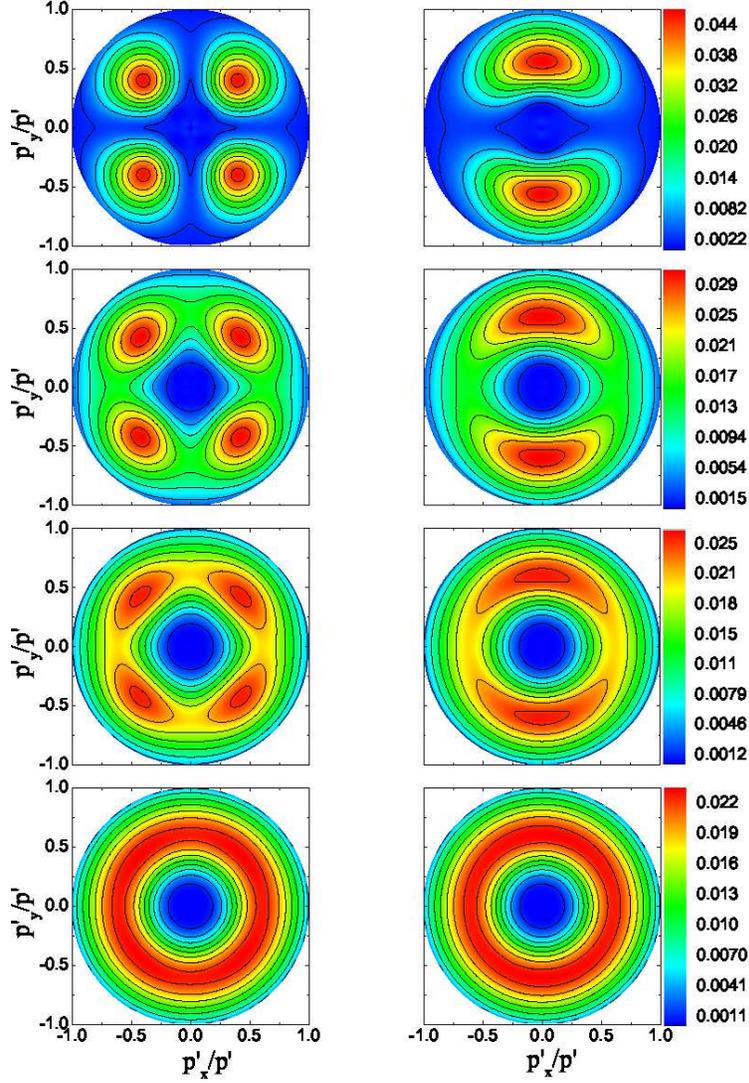}
\caption{(Color online)
Similar to Fig. {\ref{fig4}} but for  $ \omega_{3} =3\omega_1 $.
The laser field intensities are
$I_3=10I_1,I_1,10^{-1}I_1,$ and $10^{-3}I_1$ from top to bottom,
 with $I_1=10^{12}$ W/cm$^2$.
The rest of the parameters are the same as in  Fig. {\ref{fig4}}.
The magnitudes of the DCSs in a.u.  are indicated by the color scales in each row.}
\label{fig7}
\end{figure}

\begin{figure}
\centering
\includegraphics[width=5in,angle=0]{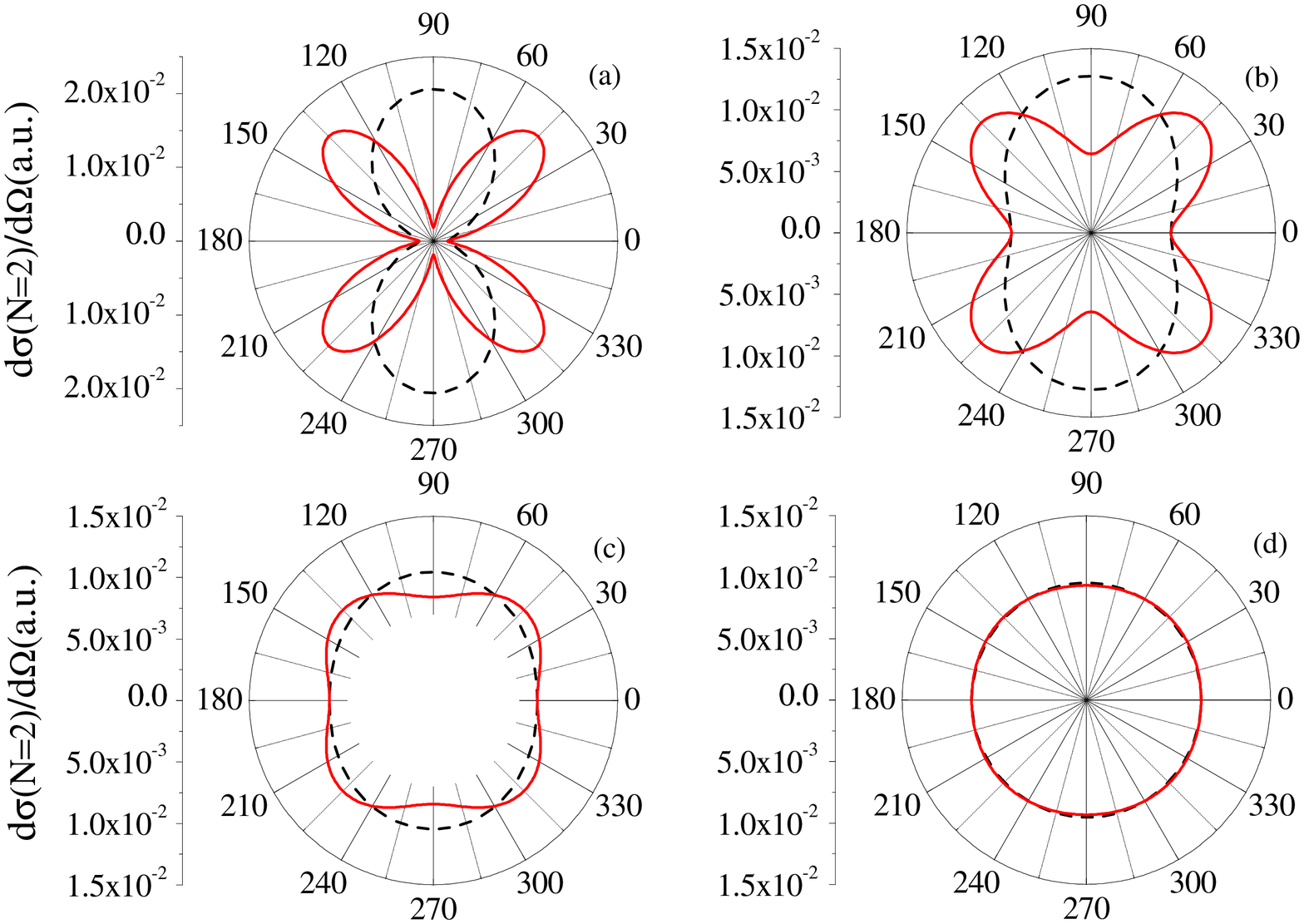}
\caption{(Color online)
Similar to Fig. {\ref{fig6}} but for $\omega_{3} =3\omega_1 $,
at $I_3=10I_1$ in panel (a), $I_1$ in panel  (b), $10^{-1}I_1$ in panel  (c),
 and $10^{-3}I_1$ in panel  (d).
 The rest of the parameters are the same as in  Fig. {\ref{fig4}}.
The DCSs by the two-color corotating CP fields are invariant to rotation by an $\pi$ angle while
by the counter-rotating CP fields are invariant to rotation by an $\pi/2$ angle.
}
\label{fig8}
\end{figure}

Figure \ref{fig7} presents similar results as in Fig. \ref{fig4} but for
a combination of the fundamental laser field and
its third harmonic, $\omega_3= 3\omega_1 $, which results
in the quiver motion amplitude $\alpha_{03} = \alpha_{01} \sqrt{f}/9$
and the argument of the Bessel function $R_3 =R_1 \sqrt{f}/9$.
In order to understand the different symmetry patterns in Fig. \ref{fig7},
 we show  DCSs in Fig. \ref{fig8}  for counter-rotating (solid lines) and corotating (dashed lines)
CP fields  as a function of the azimuthal angle $\varphi$,
 at the scattering angle $\theta = 20^\circ $, and the harmonic field intensity
$I_3= f I_1$,  with  laser intensity ratios $f =10$ in Fig. \ref{fig8}(a), $1$ in Fig. \ref{fig8}(b),
 $10^{-1}$ in Fig. \ref{fig8}(c), and $10^{-3}$ in Fig. \ref{fig8}(d).
In contrast with the $\omega_2=2\omega_1 $ case we found that the projectile electron
 is scattered with a  high probability in the directions of the azimuthal angles
$ \varphi= 45^\circ, 135^\circ,  225^\circ$, and  $ 315^\circ $
for counter-rotating CP fields.
DCS has a \textquotedblleft four-leaf clover\textquotedblright
pattern which is described in the weak field  domain,
Eqs. (\ref{t3lp})  and (\ref{tne3}), by a $ \cos 4\varphi$ term
 and is invariant  to rotation around the $z$ axis by an azimuthal angle
  $\varphi=\pi/2$ for  counter-rotating CP fields.
For corotating CP fields, DCS has a specific pattern described by $ \cos 2\varphi$
and the invariant rotation angle is  $ \varphi=\pi$.
The DCSs  for co- and counter-rotating polarizations
are symmetric with respect to reflection in the $(x,z)$- and $(y,z)$-planes, such that
${d{\sigma}_{m=3}}( \theta, \pi-\varphi )/{d\Omega}
 ={d{\sigma}_{m=3}}( \theta, \pi+\varphi )/{d\Omega}$
and
${d{\sigma}_{m=3}}( \theta, \pi/2-\varphi )/{d\Omega}
 ={d{\sigma}_{m=3}}( \theta, \pi/2+\varphi )/{d\Omega} $, respectively.
Recently,  three-dimensional electron distributions with one and three lobes were
 experimentally observed  in strong-field ionization by two-color CP fields \cite{Mancuso15},
using a superposition of fundamental and second harmonic of a Ti:sapphire laser.
Similar rotational and reflection symmetries
were obtained in above-threshold detachment of negative fluorine ions
 by a two-color bicircular laser field \cite{Odzak}.

\noindent
As expected, at relatively low harmonic intensities $I_{m}  \ll I_1$ ($m=2,3$),
 where the  one-color ($\omega_1$) two-photon processes are dominating,
  the DSCs are almost independent on the azimuthal angle $\varphi$
and have nearly circularly symmetric patterns in the polarization plane \cite{acgabi2},
 as  shown in Figs. \ref{fig4} and \ref{fig7} and
 at $\theta=20^{\circ}$ in Figs. \ref{fig6}(d) and \ref{fig8}(d) for two left-handed-CP pulses and
 two left- and right-handed-CP pulses, respectively.
In conclusion, the DCSs presented in Fig. \ref{fig4} and Figs.  \ref{fig6}-\ref{fig8}
for both co- and counter-rotating CP fields have different interference patterns
between the different paths leading to the
same final state and  have almost the same  peak magnitude.
The counter-rotating circular polarization case has a different symmetry profile
and obeys different symmetry operations:
The DCS is invariant  to rotation of the projectile electron momentum  around the $z$ axis by an angle
 $2\pi/(m+1)$  for  counter-rotating CP fields, whereas for the corotating CP fields
 the invariant rotation angle is  $2\pi/(m-1)$.

\section{Summary and conclusions}
\label{V}

Using a semiperturbative method, we have studied the electron-hydrogen scattering  by
a two-color CP laser field of commensurate photon energies $\omega_1$ and $m\omega_1$,
and derived useful analytical formulas for DCSs that are valid for
circular and/or  linear polarization and give more physical insight of the scattering process
and valuable information for experimental investigations.
A comparison between the two-photon absorption DCSs for two-color co- and counter-rotating
CP laser fields is made at different photon energies of the harmonic field
$2\omega_1$ and $3\omega_1$ ($m=2$ and $3$),
and the effect of the intensity ratio of the two-color laser field components
 on the DCSs is analyzed.
The DCSs of the scattered electrons by hydrogen atoms in a two-color bicircular
laser field of photon energies $\omega_1$ and $m\omega_1$
present a rotational symmetry  with respect to rotation of the projectile electron momentum
by an azimuthal angle $2\pi/(m-1)$ for corotating polarizations,
while for counter-rotating polarizations the invariant rotation angle is $2\pi/(m+1)$.
In addition to the rotational symmetry, for the studied  scattering geometry
$\bm{\varepsilon}_{\pm}  =(\mathbf{e}_x \pm i \mathbf{e}_y)/\sqrt{2}$
and $\bf p \parallel \mathbf{e}_z$,
  the DCSs  are symmetric with respect to reflection in the $(x,z)$ plane
for both  $m=2$ and $3$,  while the $(y,z)$ plane is
a reflection symmetry plane for $m=3$  only.
It was found that the modification of the photon helicity implies a change
in the symmetries of the DCSs and by changing the laser field intensity ratio
the angular distribution of the scattering signal can be modified.
By choosing the photon energies ratio to be even ($m=2$) or odd ($m=3$) and
varying the intensity ratio of the   co- and counter-rotating two-color CP laser field
components  we can manipulate the angular distribution of the scattered electrons.
The optimization of the scattering signal in laser-assisted electron-hydrogen
 scattering process depends for $\omega_3=3\omega_1$ on the intensity ratio
of the two-color laser field components,
whereas for $\omega_2=2\omega_1$ depends on the scattering geometry,
 fundamental field intensity, and photon energy.

\newpage
\appendix

\section{SECOND-ORDER CORRECTION TO THE ATOMIC WAVE FUNCTION FOR A TWO-COLOR LASER FIELD}
\label{A1}

In order to calculate  the  second-order atomic transition amplitude,
 Eq. (\ref{t2}),
we use the expression of the  quadratic response tensors defined in Ref. \cite{vf2}
for the two-color laser field given by Eq. (\ref{field})
to obtain the second-order correction to the atomic ground  state as
\begin{eqnarray}
  \psi_{1s}^{(2)}(\mathbf{ r},t) &=&
\sum_{k=1,m}\frac{\alpha_{0k}^2 \omega_k^2}{4}
 \sum_{j,l=1}^{3} \left[
 \varepsilon_{kj} \varepsilon_{kl}
\; w_{jl,100} ( \Omega_k^{\prime \,+}, \Omega_{k}^{+};\mathbf{r}) e^{-2i\omega_kt}
+  \varepsilon_{kj}^* \varepsilon_{kl}^*
\;w_{jl, 100}( \Omega_k^{\prime \,-}, \Omega_k^{-};\mathbf{r}) e^{2i\omega_kt}
\right.
\nonumber \\ && \left.
+\varepsilon_{kj} \varepsilon_{kl}^* \; \widetilde w_{jl,100} ( E_{1s}, \Omega_{k}^{-};\mathbf{r})
+\varepsilon_{kj}^* \varepsilon_{kl} \; \widetilde w_{jl, 100}( E_{1s}, \Omega_k^{+};\mathbf{r})
  \right]\nonumber \\&&+
\frac{\alpha_{01} \alpha_{0m}\omega_1\omega_m}{4}
 \sum_{j,l=1}^{3} \left\{
 \varepsilon_{1j} \varepsilon_{ml}
[ w_{lj,100} ( \Omega^{\prime \,+}, \Omega_{1}^{+};\mathbf{r})
+ w_{jl,100}( \Omega^{\prime \,+}, \Omega_m^{+};\mathbf{r}) ]
e^{-i(\omega_1+\omega_m)t}\right. \nonumber \\ &&
+ \varepsilon_{1j}^* \varepsilon_{ml}^*
[w_{lj, 100}( \Omega^{\prime \,-}, \Omega_1^{-};\mathbf{r})
+ w_{jl, 100}( \Omega^{\prime \,-}, \Omega_{m}^{-};\mathbf{r})]
e^{i(\omega_1+\omega_m)t}
\nonumber \\ &&
+ \varepsilon_{1j} \varepsilon_{ml}^*
[ w_{lj,100} ( \Omega^{ +}, \Omega_{1}^{+};\mathbf{r})
+ w_{jl,100}( \Omega^{ +}, \Omega_m^{-};\mathbf{r}) ]
e^{-i(\omega_1-\omega_m)t}
 \nonumber \\ && \left.
+ \varepsilon_{1j}^* \varepsilon_{ml}
[w_{lj, 100}( \Omega^{-}, \Omega_1^{-};\mathbf{r})
+ w_{jl, 100}( \Omega^{ -}, \Omega_{m}^{+};\mathbf{r})]
e^{i(\omega_1-\omega_m)t}
\right\}
,
\end{eqnarray}

\noindent
where the tensors $w_{jl, 100}$ and $\widetilde w_{jl,100}$ are defined by
Eqs. (\ref{wb2}) and (\ref{wt}), and
the energy parameters are $\Omega_k^{ \pm}=E_{1s} \pm  \omega_k $,
$\Omega_k^{\prime \, \pm}=E_{1s} \pm 2 \omega_k $,
$\Omega^{ \pm} = E_{1s} \pm ( \omega_1-\omega_m)$,
and
$\Omega^{\prime \, \pm} = E_{1s} \pm ( \omega_1 +\omega_m)$, with $k=1$ and $m$.

\section{APPROXIMATE FORMULA FOR THE GENERALIZED BESSEL FUNCTIONS
$B_N(R_1,R_m,\phi_{1}, \phi_{m})$ }
\label{A2}

We recall the definition introduced in Subsec.  \ref{scm} of the phase-dependent
 generalized Bessel function for commensurate photon energies $\omega_1$ and
$\omega_m=m\omega_1$,
\begin{equation}
 B_N(R_{1},R_{m};\phi_{1},\phi_{m})
=\sum_{l=-\infty}^{+\infty}
 J_{N-m l}(R_{1})  J_{l}(R_{m})\exp{[-i  l(m\phi_{1}- \phi_{m})]},
\label{bnn}
\end{equation}
which   is a $2\pi$ periodic function  in $ \phi_{1}/m$ and $ \phi_{m}$ \cite{varro}.
Whenever the argument of the Bessel functions of the first kind is small, i.e.,
$R_{1} \ll 1$ and $R_{m} \ll 1$, which is satisfied at low laser intensities or
at small scattering angles with moderate laser intensities,
 the approximate expressions of the generalized Bessel functions  can be used.
In the present paper we keep the second-order terms in the fields in the
transition amplitudes and, therefore, we restrict to the following approximate formula:
\begin{equation}
B_N \simeq
J_N(R_1)J_0(R_m)+  J_{N-m}(R_1)J_1(R_m)e^{-i(m\phi_{1}- \phi_{m})},
            \quad \text{for  \;  $ N \geq  0 $}.
\end{equation}
\noindent
Using the approximate relation  $ J_N(R_k)\simeq  R_k^N/(2^N {N!})$
and the symmetry formula $ J_{-N}(R_k) = (-1)^N J_{N}(R_k)$
of the Bessel functions of the first kind \cite{Watson}, we further obtain
\begin{equation}
B_N  \simeq
 \frac{R_1^N}{2^NN!} +
 \frac{ (\beta R_1)^{|N-m|}R_m}{ 2^{|N-m|+1} \; |N-m|!}e^{-i(m\phi_{1}- \phi_{m})},
 \quad \text{for  \;  $ N \geq  0  $},
\label{bnap}
\end{equation}
where
$ \beta=-1$  for   $0< N  <m  $ and $\beta=1$ for  $ N \geq  m  $.
Because the appropriate leading terms are kept in the total transition amplitude
in the weak field domain, Eq. (\ref{tep}),
the following approximate expressions are used for
 two-photon absorption scattering process:
$B_0 \simeq 1$, $B_1 \simeq R_1/2$,
$B_2 \simeq R_1^2/8  + (\beta R_1/2 )^{|2-m|}(R_m/2)e^{-i(m\phi_{1}- \phi_{m})}/|2-m|!$,
 and $B_N \simeq 0$ for $N \geq 3  $.
Explicitly, for $N=2$ the generalized Bessel function   is approximated as
\begin{equation}
B_2 \simeq
\begin{cases}
[ \alpha_{01}^2(\bm{\varepsilon}_1 \cdot \mathbf{q})^2
 + 4\alpha_{02} \,\bm{\varepsilon}_2 \cdot \mathbf{q} ] \, e^{-i2\phi_1}/8,
   &    \quad \text{for  \;  $m=2  $}, \\
[\alpha_{01}^2(\bm{\varepsilon}_1 \cdot \mathbf{q})^2
 - 2\alpha_{01}\alpha_{03} (\bm{\varepsilon}_1^* \cdot \mathbf{q})(\bm{\varepsilon}_3 \cdot \mathbf{q} ) ]
\, e^{-i2\phi_1}/8,
 &    \quad \text{for  \;  $m=3  $}.
 \end{cases}
\end{equation}

\noindent
Similar results can be obtained for $ N <  0$ by using the symmetry relation for $B_{-N}$,
$$B_{-N}(R_{1},R_{m};\phi_{1},\phi_{m}) = (-1)^N B_{N}^{\mbox *} [R_{1},R_{m};\phi_{1},\phi_{m}+(m-1)\pi].$$
For odd harmonic orders, such that $m=2s+1$, with $s$ being a positive integer, the following symmetry relation occurs:
$B_{-N}(R_{1},R_{m};\phi_{1},\phi_{m}) = (-1)^N B_{N}^{\mbox *} (R_{1},R_{m};\phi_{1},\phi_{m})$.
For even harmonic orders, such that $m=2s$, the following symmetry relation holds:
$B_{-N}(R_{1},R_{m};\phi_{1},\phi_{m}) = (-1)^N B_{N}^{\mbox *} (R_{1},R_{m};\phi_{1},\phi_{m}+\pi)$.

\section*{ACKNOWLEDGMENTS}
The work by G. Buica was supported by  research programs PN 16 47 02 02
 Contract No. 4N/2016  and  FAIR-RO Contract No. 01-FAIR/2016
from the UEFISCDI and the Ministry of Research and Innovation of Romania.

\clearpage
\newpage

\newpage

\newpage


\begin{thebibliography}{99}


\bibitem{ehl2001} F. Ehlotzky,
Phys. Rep. \textbf{345}, 175 (2001).

\bibitem{plasma}
 Y. Shima and H. Yatom,  Phys. Rev. A \textbf{12}, 2106 (1975);
 M. B. S. Lima, C. A. S. Lima, and L. C. M. Miranda, \textit{ibid.} \textbf{19}, 1796 (1979).

\bibitem{astro}
S. Chandrasekhar,
\textit{An Introduction to the Study of Stellar Structure},
 (Dover, Mineola, NY, 1967), p. 261;
M. J. Seaton, in \textit{Advances in Atomic, Molecular and Optical Physics}
edited by B. Bederson and A. Dalgarno
 (Academic Press, New York, 1994), p. 395.

\bibitem{massey}N. F.  Mott and H. S.  W. Massey,
\textit{The Theory of Atomic Collisions} (Oxford University Press, London, 1965);
C. Joachain, \textit{Quantum Collision Theory} (North Holland, Amsterdam, 1987).

\bibitem{mason}  N. J. Mason,
 Rep. Prog. Phys. \textbf{56}, 1275 (1993).

\bibitem{ehl1998} F. Ehlotzky, A. Jaro\'n, and J. Z. Kami\'nski,
Phys. Rep. \textbf{297}, 63 (1998).

\bibitem{bransden} B. H. Bransden and C. J. Joachain,
\textit{Physics of Atoms and Molecules} (Longman, London, 1983).

\bibitem{joa2012} C. J. Joachain, N. J. Kylstra, and R. M. Potvliege,
 \textit{Atoms in Intense Laser Fields}
(Cambridge University Press, Cambridge, UK, 2012), p. 466.

\bibitem{musa} M. O. Musa, A. MacDonald, L. Tidswell, J. Holmes, and B. Wallbank,
 J. Phys. B: At. Mol. Phys. \textbf{43}, 175201  (2010).

\bibitem{Kanya}
R. Kanya, Y. Morimoto, and K. Yamanouchi,
Phys. Rev. Lett. \textbf{105}, 123202 (2010).

\bibitem{Harak}
B. A. de Harak, L. Ladino, K. B. MacAdam, and N. L. S. Martin,
Phys. Rev. A \textbf{83}, 022706 (2011).

\bibitem{Kanya2}
Y. Morimoto, R. Kanya, and K. Yamanouchi,
Phys. Rev. Lett. \textbf{115}, 123201 (2015);
R. Kanya, Y. Morimoto, and K. Yamanouchi,
 in \textit{Progress in Ultrafast Intense Laser Science X},
edited by K. Yamanouchi, G. Paulus,  and D. Mathur
(Springer, Cham, Switzerland, 2014), p. 1.

\bibitem{gabi2017}  G. Buica, Phys. Rev. A {\bf 92}, 033421 (2015);
J. Quant. Spectrosc. Radiat. Transf.  \textbf{187}, 190 (2017).

\bibitem{Long}S. Long,  W. Becker, and J. K. McIver,
 Phys. Rev. A \textbf{52}, 2262 (1995).

\bibitem{Eichmann}  H. Eichmann, A. Egbert, S. Nolte, C. Momma, B. Wellegehausen, W. Becker, S. Long,  and J. K. McIver, Phys. Rev. A \textbf{51}, R3414 (1995).

\bibitem{Fleischer2014}
A. Fleischer, O. Kfir, T. Diskin, P. Sidorenko, and O. Cohen,
 Nat. Photon. \textbf{8}, 543 (2014).

\bibitem{Mancuso15}
C. A. Mancuso, D. D. Hickstein, P. Grychtol, R. Knut, O. Kfir, X. M. Tong, F. Dollar,
 D. Zusin, M. Gopalakrishnan, C. Gentry, E. Turgut, J. L. Ellis, M. C. Chen,
A. Fleischer, O. Cohen, H. C. Kapteyn, and M. M. Murnane,
 Phys. Rev. A \textbf{91}, 031402(R) (2015).

\bibitem{Mancuso16}
 C. A. Mancuso, K. M. Dorney, D. D. Hickstein, J. L. Chaloupka,
J. L. Ellis, F. J. Dollar, R. Knut, P. Grychtol, D. Zusin, C. Gentry,
M. Gopalakrishnan, H. C. Kapteyn, and M. M. Murnane,
Phys. Rev. Lett. \textbf{117}, 133201 (2016).

\bibitem{Odzak1}
S. Od\v{z}ak and D. B. Milo\v{s}evi\'{c},
Phys. Rev. A \textbf{92}, 053416 (2015).

\bibitem{Odzak}
S. Od\v{z}ak, E. Hasovi\'{c}, W. Becker, and D. B. Milo\v{s}evi\'{c},
J. Mod. Opt. \textbf{64}, 971 (2017).

\bibitem{acgabi2} A. Cionga, F. Ehlotzky, and G. Zloh,
Phys. Rev. A {\bf 61}, 063417 (2000).

\bibitem{acgabiopt} A. Cionga, F. Ehlotzky, and G. Zloh,
Phys. Rev. A {\bf 62}, 063406 (2000);
J. Phys. B: At. Mol. Opt. Phys. \textbf{33}, 4939 (2000).

\bibitem{b-j}
F.  W. Byron Jr. and C.  J. Joachain,
J. Phys. B {\bf17}, L295 (1984).

\bibitem{volkov}
D. M. Volkov,
Z. Phys. \textbf{94}, 250 (1935).

\bibitem{vf1}
V. Florescu and T. Marian,
 Phys. Rev. A {\bf 34}, 4641 (1986).

\bibitem{vf2}
 V. Florescu, A. Halasz, and M. Marinescu,
Phys. Rev. A {\bf 47}, 394 (1993).

\bibitem{Watson} G. N. Watson,
{\it Theory of Bessel Functions} (Cambridge University Press, Cambridge, UK, 1962).

\bibitem{varro} S. Varr\'{o} and F. Ehlotzky,
J. Phys. B: At. Mol. Opt. Phys. \textbf{28}, 1613 (1995).

\bibitem{bf}  F. V. Bunkin and M. V. Fedorov,
 Zh. Eksp. Teor. Fiz., \textbf{49}, 1215 (1965)
[Sov. Phys. JETP {\bf 22}, 844 (1966)].

\bibitem{milo}D. B. Milo\^sevi\'c, F. Ehlotzky, and B. Piraux,
 J. Phys. B \textbf{30}, 4347 (1997).

\bibitem{acgabi3} A. Cionga, F. Ehlotzky, and G. Zloh,
Phys. Rev. A {\bf 64}, 043401 (2001).

\bibitem{gavrila1970}M. Gavrila and A. Costescu,
Phys. Rev. A \textbf{2}, 1752 (1970).

\bibitem{gavrila}
V. Veniard, M. Gavrila, and A. Maquet,
Phys. Rev. A \textbf{35}, 448(R) (1987);
A. A. Krylovetskiĭ, N. L. Manakov, S.  I. Marmo, and A. F. Starace,
J. Exp. Theor. Phys.  \textbf{95},  1006 (2002).

\bibitem{manakov}
N. L. Manakov, A. V. Meremianin, J.  P.  J. Carney, and R. H. Pratt,
 Phys. Rev. A \textbf{61}, 032711 (2000).

\bibitem{taieb} R. Taieb, V. Veniard, A. Maquet, N. L. Manakov, and S. I. Marmo,
Phys. Rev. A \textbf{62}, 013402 (2000).

\bibitem{fifirig2000}
M. Fiﬁrig, V. Florescu, A. Maquet, and R. Ta\"{\i}eb,
J. Phys. B: At. Mol. Opt. Phys. \textbf{33}, 5313 (2000);
M. Fiﬁrig, A. Cionga, and F. Ehlotzky,
Eur. Phys. J. D \textbf{23}, 333 (2003).

\bibitem{starace}A. Y. Istomin, E. A. Pronin, N. L. Manakov, S.  I. Marmo, and A. F. Starace,
Phys. Rev. Lett. \textbf{97}, 123002 (2006).

\bibitem{acgabi99} A. Cionga and G. Zloh,
  Laser Phys. {\bf 9}, 69  (1999).

\end{thebibliography}
\end{document}